\documentclass[5p,times]{elsarticle}

\usepackage{amssymb}
\usepackage[numbers]{natbib}
\usepackage{multicol}
\usepackage{multirow}
\usepackage{clrscode3e}
\usepackage{tabu}
\usepackage{algorithm}
\usepackage{algpseudocode}
\usepackage{listings}

\biboptions{sort&compress}

\usepackage{url}

\usepackage{breakurl}
\usepackage[breaklinks]{hyperref}

\journal{Journal of Network and Computer Applications}

\begin{document}

\begin{frontmatter}



\title{Making Tax Smart: Feasibility of Distributed Ledger Technology for building tax compliance functionality to Central Bank Digital Currency}


\author[inst1]{Panos Louvieris\corref{cor}}
\ead{panos.louvieris@brunel.ac.uk}
\author[inst1]{Georgios Ioannou}
\ead{georgios.ioannou@brunel.ac.uk}
\author[inst1]{Gareth White}
\ead{gareth.white@brunel.ac.uk}

\cortext[cor]{Corresponding author}

\affiliation[inst1]{organization={Department of Computer Science, College of Engineering, Design and Physical Sciences, Brunel University London},
            addressline={Kingston Lane}, 
            city={Uxbridge},
            postcode={UB8 3PH}, 
            state={Middlesex},
            country={United Kingdom}}

\begin{abstract}
The latest advancements in Distributed Ledger Technology (DLT), and payment architectures such as the UK's New Payments Architecture, present opportunities for leveraging the hidden informational value and intelligence within payments. In this paper, we present Smart Money, an infrastructure capability for a Central Bank Digital Currency (CBDC) which enables real-time Value Added Tax split payments, oversight, controlled access and smart policy implementation. This capability is implemented as a prototype, called Making Tax Smart (MTS), which is based on the open source R3 Corda framework. The results presented herein confirm that it is feasible to build a MTS capability which is scalable and co-exists with the current payment systems. Smart Money CBDC has the potential to mobilise payments data in order to transform the role of money from a blunt instrument to a government policy sensor and actuator without disrupting the existing money system. DLT, smart contracts and programmable money have a crucial role to play with benefits for government departments, the economy and society as a whole.
\end{abstract}




\end{frontmatter}


\section{Introduction}
\label{introduction}
Distributed Ledger Technology (DLT) is a major catalyst for change in the money system and the administration and control of money.  Furthermore, DLT offers promising potential for disrupting the existing information management frameworks of government departments, such as the BoE and the HMRC in order to overcome the friction of public policy decision making. DLT’s are already being employed in diverse business applications and have matured significantly since the inception of the first established network of this kind, Bitcoin \cite{Nakamoto2008}. This is the first and most popular digital currency and payment system employed on a DLT platform \cite{Hayes20171308} and was the first of its kind to eliminate the need for a single entity that accounts for and controls a currency \cite{Eyal201738}. Its success has led to a market capitalisation that peaked \$1tn in April 2021 and the subsequent emergence of numerous alternative digital currencies (also known as cryptocurrencies or cryptoassets) such as LiteCoin, Ethereum, Ripple and Bitcoin Cash.

The success of digital currencies is attributed to the perceived failures of governments and central banks during the 2008 financial crisis including serving as cheaper alternatives to existing debit and credit card systems \cite{Fry2016343}. Just as cash, i.e. banknotes, digital currencies are considered as medium of exchange, store of value and units of account \cite{AMMOUS2018}. However, the adoption of de-regulated cryptocurrencies creates problems associated with: the volatility of cryptoassets which pose a risk for buyers and sellers, the risk of payments systems which operate outside the regulatory framework and the use of cryptocurrencies for money laundering and financing terrorist groups \cite{VANDEZANDE2017341}.

Meanwhile, the role of DLT has evolved from a "digital currency ledger" to a medium for realising transactions between unknown participants \cite{Dai201822970}. Further applications include disintermediated environments that enable trust-less payments \cite{HAWLITSCHEK201850} and decentralised and immutable storage \cite{YEOW2017}. DLT is used in a number of government applications. Examples include notary services for citizens \cite{SULLIVAN2017470}, energy trading \cite{Muzumdar2021}, securing archives of public documents \cite{Collomosse2018}, B2G information sharing \cite{Engelenburg2019} and land registry \cite{Olnes2017355}. Central banks across the globe, as part of public sector explorations of DLT, are considering providing DLT-enabled digital currencies \cite{Koning2016,Bech2017,DeutscheBundesbank2017} directly to the general public through accounts held in their books \cite{Engert2017} as a complement to fiat currency which is currently offered via cash or through commercial banks. Recent research efforts around the globe lead by central banks \cite{Qian2019,BankofCanada2020,Riksbank2021,England2020} actively explore the characteristics, design criteria, implications and acceptance of Central Bank Digital Currencies also known as CBDC’s. Developments in DLT, particularly those related with expanding the functionality of smart contracts, accentuate the concept of 'programmable money' \cite{England2020} which creates opportunities for fiscal policy implementation. Similarly, controlling spending on social benefits such as Universal Credit \cite{Royston2012} can be provided in the form of special purpose tokens which can be spent exclusively for particular classes of goods and services e.g. as food, rent and medicine. A CBDC can operate as a public policy actuator with just-in-time targeted interventions in the event of financial or health crises with large economic impact.

The advent of DLT and CBDC comes at a time where payments are undergoing radical changes in a global scale, triggered by events such as the implementation of the Payment Services Directive 2 \cite{Romanova2018}, Open Banking \cite{Wang2019a}, the ISO 20022 electronic data interchange standard \cite{InternationalOrganisationforStandardization2020}, the transformation of payment systems such as UK's New Payments Architecture (NPA) \cite{Pay.UK} and the transformation of cross-border payments via the introduction of SWIFT GPI \cite{SWIFT2022}. Payment messages are now carrying "enhanced data" elements concerned with contextual information of a transaction, e.g. item description, item quantity, unit cost, VAT invoice number, VAT amount, Legal Entity Identifier and more. This poses an important question concerning how this data should be exploited for its information value within a proposed Smart Money system for greater societal and economic benefit?

We present Smart Money, a Central Bank Digital Currency Architecture as an essential component for an advanced digital economy. This work is motivated by the Bank of England's desire to exploit CBDCs as a potential and feasibile supplement for cash which has implications for the payment system and government authorities \cite{England2021}. Furthermore, this paper considers that CBDCs have a additional role, which is to assist Other Government Departments' (OGDs) policies such as automating tax payments at the point-of-sale. For the first time, the design and implementation of a feasible DLT-based approach which integrates smart contracts with the existing and future payment infrastructure and causes minimal intervention and disruption to the financial system is proposed in this paper. The design specification proposed herein, facilitates trusted information management and sharing among OGDs in order to mobilise data and introduce smart policy making. 

Of particular interest is the issue of VAT split payments, where the VAT is deducted from a payment to be transferred directly to His Majesty's Revenue and Customs (HMRC) and the supplier receives the net amount of the transaction. Within the Making Tax Smart prototype presented herein, split payments are processed via smart contracts based on the information recorded on the invoice. This approach leverages the informational value offered by payments enhanced data feature, presented in the payment messages of the NPA \cite{Pay.UK}, providing further benefits for OGDs; specifically, CBDC facilitates gathering of economic relevant data in near real-time to support compliance and oversight of public policy planning and development. Lastly, MTS introduces 'smart warrants', a mechanism that enables permissioned access control over the transaction data. 
\subsection{Aim}
\label{aim}
Develop a CBDC which provides a real-time precision information management capability to transform money by enhancing its informational value for government policy applications in a feasible and scalable manner.

\subsection{Objectives}
\label{objectives}
\begin{enumerate}
\item Review the current efforts in developing DLT-enabled CBDCs and the opportunities that this technology brings for overcoming the frictions of government policy making.
\item Design a DLT-based CBDC architecture that provides new capabilities for governments which are non-disruptive to existing payment systems. The objectives of these capabilities are: i) To utilise smart contracts for delivering near real-time tax collection via VAT split payments and permissioned data access control for tax compliance via smart warrants and ii) to demonstrate the programmable money concept for implementing smart policies to ensure control over spending.
\item Demonstrate the feasibility  and scalability of the CBDC architecture within the context of a smart contract-enabled VAT split payments use case.
\end{enumerate}

\subsection{Research contributions}
\label{contributions}
Through the exploration of a Central Bank Digital Currency, this paper will show that the traditional definition of money (medium of exchange, store of value, unit of account) is limited. Due to the latest advancements in Distributed Ledger Technology and the increasing demand for digital government transformation, money can now be a significant catalyst for change by extending the traditional definition of money to include "money as a sensor and a policy actuator". Smart Money is the implementation of this new definition of money, and this paper demonstrates the feasibility and scalability of the Smart Money proof-of-concept within the context of the VAT split payment use case presented here.

The rest of the paper is as follows: Section 2 covers the current state-of-the-art in CBDCs and the issues that this tech promises to resolve which concern state governments. In Section 3 the motivation behind Smart Money and the intersection of DLT-enabled capabilities with smart policies is presented. Section 4 provides the specification of the Smart Money system and the methodology for evaluating the feasibility and scalability of the design are presented in Section 5. Sections 6 and 7 present the results of the feasibility and scalability tests. Finally, the conclusions are presented in Section 8.

\section{Related Work}
\subsection{Distributed Ledger Technology Frameworks}
A distributed ledger constitutes a medium of transaction storage for participating entities or nodes whereby all participants maintain a synchronised copy of the ledger. The rules for accessing the ledger are determined by the characteristics and requirements of the DLT application. Trust-less ecosystems such as the Bitcoin and Ethereum blockchains utilise a public ledger \cite{RAHMAN2018171} where all transaction data is publicly available albeit user identities are encrypted. These ledgers are permissionless \cite{Sompolinsky201846} hence anyone can submit and verify transactions on the network and all network participants have the same permissions \cite{Paech2017}.
Blockchains such as the R3 Corda \cite{Khan201729} and Hyperledger Fabric \cite{Vukolic20173} which are optimised for fintech applications employ a permissioned approach. In this paradigm, the ledger is accessible only by authenticated entities and the level of access for each entity is defined by the business logic \cite{Eyal201738}. For example, customers are only allowed to submit transactions (i.e. issue payments) whereas payment settlement is processed by entities such as notary services \cite{Khan201729} or delegated validating groups of servers \cite{Armknecht2015}. The permissioned approach is commonly employed in the financial sector because transactions involve personal data. Therefore, the ledgers are usually private and non-authenticated entities are not allocated any access rights. In addition, a ledger may be public and permissioned hence its contents may be publicly available but transaction submission and verification is only permitted by designated entities \cite{YEOW2017}. A special case of a permissioned access model is the consortium blockchain \cite{Niranjanamurthy20181}, where a limited number of entities is given certain privileges to validate transactions \cite{Sankar2017} . These consortiums are formed by multi-lateral agreements between businesses such as financial institutions \cite{Armknecht2015} or mobile telecommunications providers  \cite{Lee20172274}. 

\subsection{Central Bank Digital Currencies}
In addition to providing security for financial services, DLT also provides provision for secure Central Bank Digital Currencies \cite{AGGARWAL201913}. CBDC’s can be considered simply as fiat currencies issued by a central bank in place of banknotes \cite{Dyson2016, Grym2017, SverigesRiksbank2017} or complementary to cash \cite{Prasad2018}, and  are made available in electronic form through direct accounts with the central banks \cite{Ahmat2017}. Similarly, CBDC’s can be considered as an electronic form of central bank money that allow peer-to-peer transactions without the need for an intermediary \cite{Bech2017}. In addition to serving as a means-of-payment, CBDC’s are considered to operate as a store-of-value \cite{Meaning2018}. In \cite{Barrdear2016}, the term CBDC refers to “a central bank granting universal, electronic, 24x7, national-currency denominated and interest-bearing access to its balance sheet”.

So far, only two countries (Bahamas and Jamaica) have CBDCs in operation \cite{Alfonso2022}. Similar efforts are underway, where CBDC is considered as part of a new interbank settlement system \cite{Bech2017} or as a replacement for cash \cite{Riksbank2021}. The European Central Bank has also been exploring technical solutions for a CBDC, focusing on the issues of anonymity and privacy \cite{ECB2019, EuropeanCentralBank2021}. The Bank of England and HM Treasury have recently announced the joint creation of a taskforce to coordinate the exploration of a potential domestic UK CBDC \cite{BankofEngland2021}. Project Rosalind \cite{BISInnovationHub2022}, a collaboration between BOE and the Bank for International Settlements (BIS) investigates how a CBDC wallet and API for a DLT-based CBDC should be designed an implemented. At the same time, central banks are researching CBDC’s beyond their jurisdictions to explore their potential on cross-border payments. Notable examples are Project Icebreaker with the participation of the central banks of Israel, Norway and Sweden \cite{BISInnovationHub2022b} and Project mBridge which involves central banks of Hong-Kong, Thailand, UAE and China \cite{MBridge2022}.

As common and standardised definitions and frameworks for CBDCs remain to be finalised, current CBDC models are specified according to two key parameters: accessibility (wholesale versus retail) and utility (retail payments versus interbank settlements). Accessibility and utility are tightly coupled because the purpose of a CBDC prescribes who has access to the currency, namely central banks, commercial banks, “narrow banks”, non-fintech firms and households. Wholesale CBDC’s are central bank liabilities that are only accessible by banks and non-bank financial institutions. They are concerned with transactions that are processed through the Real-Time Gross Settlement (RTGS) system of a central bank which involves interbank settlements, liquidity provision and monetary policy implementation \cite{BankofEngland2016}. Retail CBDC’s are central-bank liabilities which are effectively identical to banknotes and central bank deposits and can be used as an instrument for retail payments \cite{Grym2017}. All transactions between accounts would be validated and processed by the central bank \cite{Ahmat2017} or directly from payer to payee \cite{Bordo2017}. No third-party intermediaries are involved in payments therefore transactions are settled within a settlement platform which is independent from the RTGS system \cite{SverigesRiksbank2017}. Hybrid CBDC’s aim to bridge the tradeoffs between wholesale and retail models by providing intermediary access to CDBC’s for households and non-fintech firms. This is achieved by i) exchange services provided by banks or “narrow banks” for buying and selling CBDC’s in exchange for deposits \cite{Dyson2016, Kumhof2018} or ii) by fintech firms that offer financial assets which are backed by CBDC’s \cite{Kumhof2018}. Cross-border CBDC designs \cite{MBridge2022} are also being investigated in efforts to overcome the intricacies of costly and slow correspondent-banking chains \cite{CommitteeonPaymentsandMarketInfrastructures2021}.

Outside central bank initiatives, only a handful of other, yet partial, CBDC prototypes can be found in the literature. A design for CBDCs which is based on a custom blockchain framework is proposed in \cite{Bhawana2021} but an implementation has not been provided. A demo implementation of a CBDC based on the Cosmos blockchain is presented in \cite{Han2021}, although it has not been tested for feasibility and scalability. The work of \cite{Zhang2021} proposes an implementation which is based on a custom blockchain framework using a modified Byzantine Fault Tolerance consensus scheme, however no smart contract functionality was adopted. These prototypes have not been developed to their logical conclusion neither have they been adequately tested.
The work presented herein follows a model for CBDC which is similar to Riksbank's \cite{Riksbank2021} and the R3 Corda ledger because of its unique permission and access-control management (for more details, see Section 4).

\subsection{CBDCs as programmable money}
Central banks work closely with other government departments, for example in the UK, the Bank of England works with HM Treasury and HMRC, where CBDCs have a new and major role to play. Namely, CBDCs have heralded the advent of programmable money enabling controlled policy implementation for these departments and overall governance improvement.

Conventional money has been hitherto issued, stored and spent in digital form but has only served as a blunt instrument strictly corresponding to value. Developments in distributed ledger technology have the potential to add an important feature to money thus making it 'programmable' \cite{England2020}. This property enables a number of new capabilities such as payments conditioned on rules or events \cite{Elsden2019}, delivery versus payment (DvP) settlements \cite{England2020} or auto-pay bills formalised into multi-party payment contracts in order to permit net instead of gross money flows \cite{Swan2017}.

Central banks' interest in programmable money is emerging and causes them to rethink the interface to money \cite{Ali2019}. Unfortunately. this interest remains mostly constrained to the potential for regulating cryptocurrencies \cite{IBMInstituteforBusinessValue2019} or for micropayments which are moot points and overlook the potential of programmable money for improving governance and policy implementation.

The functionality of smart contracts, which is to run generalised programs on a shared blockchain \cite{Ali2019} creates new capabilities for central banks and governments by utilising programmable money as a multiplier of public policies e.g.  monetary, financial and fiscal policies. This brings opportunities to transform the traditional role of money and the way that public policy is informed and applied. With DLT-enabled payments, purchases of goods and services can trigger VAT split payments to the tax authority so that sales tax is paid directly.

The concept of DLT-enabled real-time tax payments was put forward by \cite{Rikken2017} as a means for simplifying accounting and reducing administrative burden for businesses. Neither a practical implementation nor a design were however included. \cite{Sogaard2021} has proposed a VAT settlement solution where the entities involved (seller, buyer, banks, tax authority, auditor) are participating in an Ethereum-based DLT network. In the evaluation, the proposed software artifact does not transfer funds between any accounts i.e. seller, buyer or tax authority, neither does it confirm any account balances during the transaction. It serves exclusively as an accounting system but does not initiate or process payments. Transaction information is publicly accessible which poses confidentiality risks for the counterparties. Furthermore, the software artifact has not been empirically tested for scalability. In terms of enforcing tax compliance, the proposed artifact poses certain risks: VAT settlements are validated by an ‘auditor’ instead of the tax authority directly, which is not a proper assurance mechanism for tax compliance; also, the VAT payment is made by the seller instead of the buyer. This may cause the payment to fail if the seller is overdrawn and the balance is insufficient for fulfilling the tax payment.” Therefore, the works of \cite{Rikken2017, Sogaard2021} cannot be considered as 'programmable money' applications for VAT split payments.

In addition, the government can issue special purpose 'money' within the context of social benefits to be used for particular classes of goods and ensure that public funds are spent for the benefit of the polity. Further, programmable money can be tracked and traced facilitating 'follow-the-money' investigations on behalf of law enforcement in the pursuit for money laundering and terrorism financing. It is evident that the 'programmability' property assigns new functionality to money, which can now operate as an information sensor and a policy actuator that enables governments to utilise money to develop and enforce policy concerned with tax collection and social care.

\subsection{The opportunity of CBDCs for governance of money flows}
A DLT-enabled CBDC can provide real-time access to transactional data \cite{Campbell-Verduyn2018283} which allows larger sample sizes that support more detailed analysis regarding space and time, and is better for measuring certain behaviours by overcoming biases reported in surveys \cite{Callegaro2018}. Moreover, transactional data that is updated frequently or continuously \cite{Whitaker2017} enables an increase in sample size and granularity and brings considerable benefits beyond existing surveys for creating real-time maps of financial and activity flows across the economy \cite{Haldane2018}. These maps will enable government oversight i.e. to observe the effects of policy decisions, observe the flows of funds within the economy, monitor compliance of regulated firms \cite{Mihaylov2016} and detect payment fraud. This brings additional implications in terms of overcoming the "data siloing” effect by unlocking the transactional data stored on the ledger to OGDs and provide them with intelligence of wider coverage \cite{Adams2015, BankforInternationalSettlements2017, Callegaro2018}. Participants in a distributed ledger network can leverage the shared infrastructure to effectively streamline inter-departmental business processes and have a consistent view of the data \cite{Hileman2017}. By having a global view of the data, government departments will now have the ability to access information owned by other agencies more quickly and more transparently, respecting account holders' privacy. This has further benefits for Open Banking services \cite{Wang2019a}  whereby third-party service providers will have controlled access to account holders’ transactional data. The Open Banking route constitutes an optimal approach for integrating DLT-enabled CBDC’s with the existing payment rails, to avoid disruption to the core banking system.

\section{Smart Money}
Smart Money is an envisioned national network where UK residents possess Central Bank Digital Currencies in addition to, or in lieu of banknotes and coins.  The CBDC is not kept on a physical device such as a smartphone or a special purpose device. Instead the currency is held in special-purpose CBDC accounts which are provided for by private entities such as banks and other Payment Service Providers (PSPs). CBDC accounts can be registered by natural persons and legal entities which are permitted to hold deposits in the UK and account holders can interface with the provided services by means of electronic devices such as PC’s and mobile devices. The CBDC money flows are stored and maintained on the Smart Money Distributed Ledger which is governed by a number of departments such as His Majesty's Treasury, Bank of England, His Majesty's Revenue and Customs and Ministry of Justice. These departments have permission to access the transaction data from the DL to fulfill their policy objectives and enforce compliance in accordance with legislation which is codified in smart contracts. The Smart Money CBDC is designed to integrate with the existing and future iterations of the UK's payment system and is therefore not designed as a public cryptocurrency such as Bitcoin or Ethereum.

\begin{figure*}[!ht]
	\includegraphics[width=\textwidth]{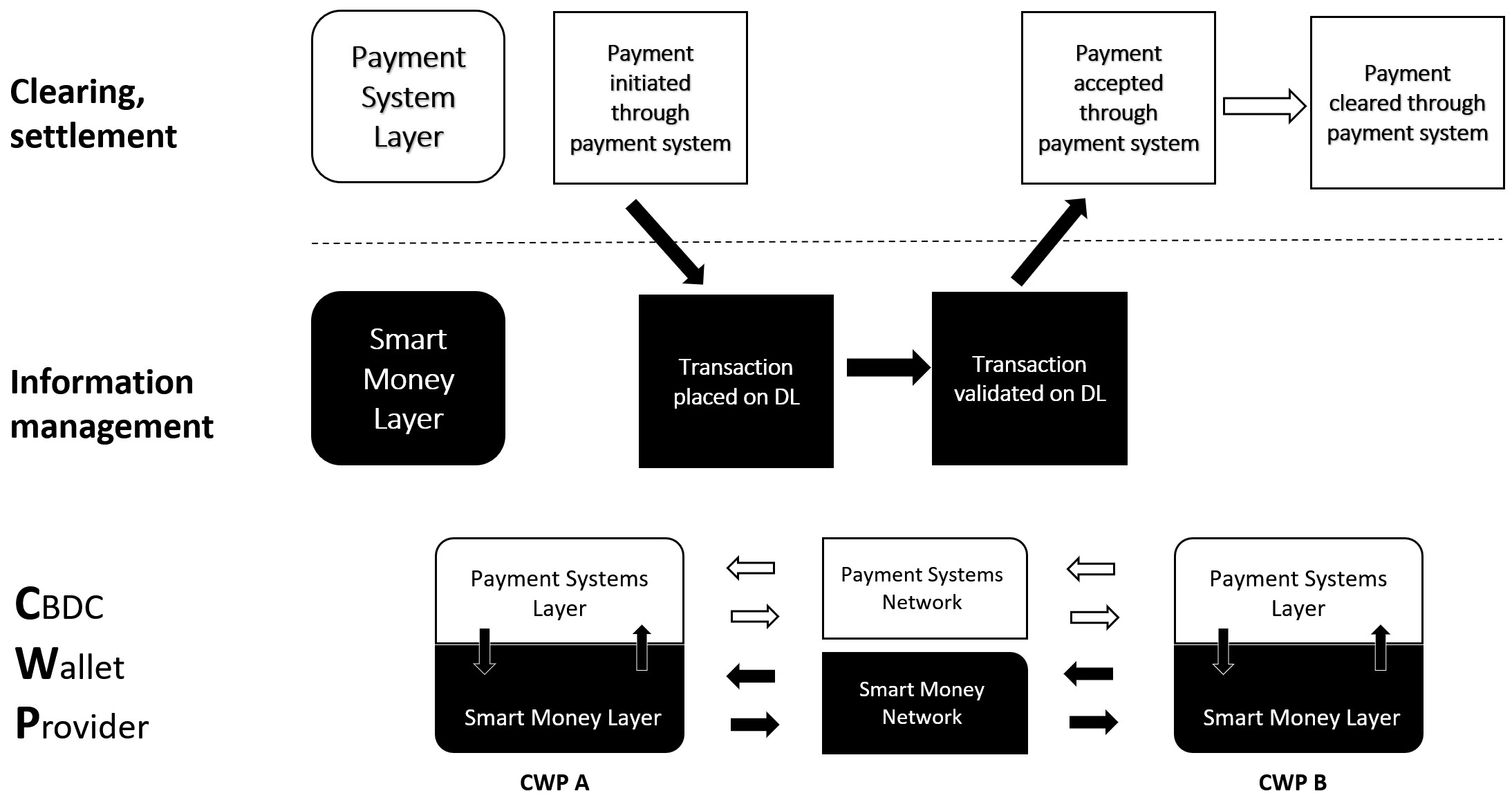}
	\caption{\label{Fig.1} The Smart Money DLT Architecture}
\end{figure*}
\subsection{Smart Money CBDC within the New Payments Architecture}
At the core of the Smart Money ecosystem sits a private and permissioned Distributed Ledger network. Financial institutions (e.g. Banks, payment service providers), government departments (e.g. Bank of England, HMRC, ONS) are direct participants of the Smart Money system, by means of dedicated nodes. In addition to payment and banking services through traditional channels, financial institutions provide CBDC wallets to their customers. Within the Smart Money system, these institutions are called CBDC Wallet Providers (CWPs). In the payments network, the CWPs continue to offer their services as in their traditional role of PSPs. In the Smart Money system, they provide CBDC payment services but also serve as data gatekeepers of their customers' payments data. This data then becomes available to government departments that are pre-authorised, such as HMRC for automating tax collection.

Financial institutions that offer payment services can operate under one of the following regimes:
\begin{enumerate}
\item Operate exclusively at the payments network and interact with the Smart Money system through a correspondent CWP.
\item Operate exclusively at the Smart Money system and interact with the payments network through a correspondent PSP.
\item Operate at both networks with a dual PSP/CWP role.
\end{enumerate}

The Smart Money system is designed to operate as an overlay to the standard payments network and records all the payments on a distributed ledger which is shared among its participants. The CWPs that participate in both the Distributed Ledger and payment networks serve as bridges between the two environments. When a CBDC wallet holder (equivalent to an account holder in the traditional banking sense) issues a payment through the payments network irrespective of the selected money instrument (CBDC or bank money), their corresponding CWP copies the message from the payment network to the Smart Money system, acting as a bridge between the two layers.

The integration of Smart Money with the payments network becomes more relevant in the wake of the forthcoming change to the UK's payment system called New Payments Architecture. The NPA adopts the ISO20022 messaging standard and introduces the following two key features of great importance to governance and oversight:
1)	Enhanced data: payment messages will now carry additional information related to utility bills, invoices, salary payments and financial assets in a standardised, structured format. 
2)	Request-to-pay: billers will be able to request payments from their customers through the payment system instead of sending an invoice or bill through conventional channels (e.g. e-mail, post). 
These novel features will enable suppliers to input granular data associated with the transaction within the payment message and transmit that information directly to their customers who can respond to that message and fulfil the payment request. This process creates a money trail of that particular transaction which matches the payment with its associated metadata. This particular money trail is replicated on the Smart Money system and stored on its distributed ledger, as part of the payment's lifecycle.

\subsection{Smart Money for enhanced oversight}
Aggregating all the money flows with enhanced data from the payment system in a common medium creates a database of unprecedented intelligence value for government departments. Payments are copied in real-time on the distributed ledger to enhance timeliness in accessing, processing and analysing payments data. Maintaining a copy of payments from across the economy (retail, wholesale, cross-border transactions) increases the coverage of intelligence which is recorded. The inclusion of enhanced data in the ledger ensures access to granular information about payments which captures their context and purpose. Moreover, government departments have smart contract-enabled access to an up-to-date database of information where integrity and accuracy are assured, thereby enhancing the availability of information.

The capability to create unparalleled intelligence value from payments transforms oversight and compliance exercised by government departments. Processes which are currently largely based on surveys and forms are now streamlined on the distributed ledger with benefits for all involved parties. Tax authorities can calculate income tax and tackle tax evasion and fraud by reviewing accurate and immutable records of payments. National statistics authorities will be able to collect unbiased and real-time data to produce high quality statistics and insights to support government decision making such as monetary and fiscal policies.

\subsection{Smart Money for Smart Policy}
Money flows are involved in many aspects of governance, such as public expenditure, taxation, financial stability, price stability or social benefits.  Unfortunately, money in its current form is a blunt instrument and does not provide the means to gather data in real-time across a broad range of economic activities. Smart Money CBDC transforms public policy by means of exercising more power on the flows of money. This is achieved by realisation of 'programmable money' through the use of distributed ledger technology and smart contracts. Specifically, sales incur Value Added Tax (VAT) and businesses are liable to pay that tax to the government. In practice, businesses submit quarterly VAT returns where the aggregate VAT for a financial term is declared. This process incurs significant administrative workload both for businesses and government, in addition to foregoing the collection of valuable transaction data. With Smart Money CBDC, VAT can be calculated and deducted at the point of transaction with the use of smart contracts. These contracts are executed by all the interested parties, i.e. the buyer, the seller and the tax authority to ensure that only the correct amount of tax can be collected. This VAT split payment feature would be impossible to carry out in a secure, transparent and trusted manner without a DL-enabled CBDC which carries enhanced data matching the payment it is spent on.

\subsection{Controlled access through Smart Warrants}
Transaction data are stored securely on the ledger with confidentiality, integrity and availability being preserved due to the properties of Distributed Ledger Technology. However, law enforcement agencies may wish to obtain access to private data when certain circumstances are present, i.e. there is probable cause that an a account holder (the data subject) has committed a criminal offence where their transaction history is highly likely to contain evidence to support a successful prosecution. Smart Money employs smart contracts for implementing functionality equivalent to conventional warrants for searching suspects’ premises albeit on payments data stored on the distributed ledger. This is achieved by incorporating a legal entity (such as a magistrate or public prosecutor) in the Smart Money system by means of a dedicated node. The purpose of that node is to authorise data access requests by investigatory authorities that are participating on the network. 

When a government department, such as HMRC launches an investigation to seek evidence of illegal behaviour within the payments data of a particular CBDC wallet holder, a Smart Warrant request is executed as part of a smart contract. The legal entity then authorises that request and grants access to the transaction history of the data subject and stores that authorisation on the ledger. HMRC can then execute the smart warrant to gain one-time access to the data subject’s account data by querying the transaction history from the corresponding CWP’s transaction ledger. To ensure separation of duties within government departments, the HMRC node hosts two separate accounts for its purposes: one account is used for authorising VAT split payments and another account is used for investigation purposes; the latter has no visibility on payments data unless a matching Smart Warrant is issued. To ensure separation of duties within government departments, the HMRC node hosts two separate accounts for its purposes: one account is used for authorising VAT split payments and another account is used for investigation purposes; the latter has no visibility on payments data unless a matching warrant is issued.

\subsection{Extending the definition of money with Distributed Ledger Technology}
With these new capabilities in place, money exceeds its traditional role as a medium of exchange, unit of account and store of value \cite{Lober2018}. Insofar, governments are employing money singly as an instrument for enforcing policy decisions within the context of the associated monetary value. To retrieve economic intelligence, governments are hitherto employing conventional, inert tools which cause frictions in the way that policy is informed and exercised. Government departments will have real-time access to all economic activity routed through the domestic payments network to obtain valuable intelligence and to control money flows in order to enforce compliance. This new role of money is accomplished by the unique approach embodied within Smart Money, which is to leverage the informational value concealed in money flows by mobilising the enhanced data and intelligence contained within the payment messages. This approach is underpinned by the distinctive features of distributed ledger technology which ensure integrity, accountability and transparency of all activity which is concerned with money flows and managing the information associated with payments. Smart Money contribution is premised on extending the traditional role of money, which now also serves as an information sensor and public policy.

\subsection{Use Case: Making Tax Smart}
Making Tax Smart (MTS) is the core use case for Smart Money, which investigates the feasibility of issuing VAT split payments for supplies of goods and services within the UK. With MTS, every payment which incurs VAT, such as purchasing a computer, paying for rent or legal services, triggers an additional payment which secures the matching value added tax for the tax agency, Her Majesty’s Revenue and Customs. The split payments are issued on the Smart Money DLT by means of smart contracts and are validated by the associated parties (i.e. Buyer, Seller, HMRC) in a VAT-able payment, namely the supplier, the customer and the HMRC. The exact amount of VAT is derived at the point of transaction by inspecting the enhanced data elements contained on the payment message i.e. the invoice issued by the supplier codified in a structured format. Split payments means the VAT from each transaction is collected by HMRC in real-time, which minimises the opportunities for fraudsters to perform VAT fraud or evade paying tax. This is of great importance to the UK tax authorities as the estimated current VAT gap (i.e. total theoretical VAT liability minus the actual VAT collected by the tax authorities) is measured at £9.8bn which amounts to 7.5\% of the theoretical VAT liability \cite{HMRC2021}. 

\section{Smart Money Corda Implementation}
To assess the scalability and feasibility of the proposed CBDC, an experimental environment for the Smart Money system has been implemented in the R3 Corda open-source blockchain platform. Corda, as a permissioned blokchain framework which has been designed for financial services, incorporates features such as accounts, tokens, notarisation services and permissioned, smart-contract enabled on-ledger access, which are suitable for the implementation of the Smart Money building blocks.

\subsection{Smart Money DLT Architecture}
The Smart Money DLT is designed to be interoperable with existing and future payment services in the UK and to not interfere with the clearing and settlement systems. The Smart Money Layer (which hosts the distributed ledger) co-exists with the Payment Systems Layer (the existing payment systems infrastructure) and stores the exchanged messages in a transaction ledger which facilitates secure and accountable exchange of information. Effectively, the DL is not used for any type of clearing/settlement processes. Its purposes are: i) to maintain an immutable registry of money transfers for rapid retrieval and analysis, ii) to facilitate citizens in exercising control and leveraging the value of their personal data and iii) to provide an audit trail for accountable data processing that preserves the privacy rights of data subjects.

In the Smart Money DLT architecture, payments are initialised by the seller by means of a "Request-to-Pay" message, or R2P in short . The recipient of the payment is responsible for registering the "enhanced data" information such as value, VAT, buyer reference, seller reference and the invoice or receipt for the transaction. In contrast to the current, paper-based scheme where contextual data concerned with the transaction is not digitally recorded, the R2P message records all the relevant information as it is routed through the payments network. The buyer's PSP receives the message from the payments network but does not take any further action until such instruction is provided by the Smart Money layer. The payment service provider of the seller is responsible for dispatching the contents of the payment message to the Smart Money layer by registering the transaction on the distributed ledger. The buyer's CWP is therefore informed for the pending transaction from both sources: the payments network and the DL. The payment however can only be authorised from the Smart Money DLT, by means of executing a smart contract. There are three stakeholders in this transaction: the buyer's CWP, the seller's CWP and the DL node of the tax agency, HRMC. HMRC's role in the payment is to enforce tax compliance by validating that the buyer allocates the correct amounts for the seller and for paying VAT. As soon as the payment has been fulfilled in the Smart Money DLT, the buyer's PSP can release the funds through the payment network and finalise the payment. The exchange of information that takes place between the two layers during the lifecycle of a payment is illustrated in Figure \ref{Fig.1}.

\subsection{Implementing Smart Money with R3 Corda}
R3 Corda is a permissioned and private distributed ledger technology platform, specially designed for financial industry applications and with smart contract support \cite{Valenta2017}. In contrast to other DLT frameworks, transactions in Corda are only announced to the involved parties on a need-to-know basis. All communications in a Corda network take the form of small multiparty protocols, called flows \cite{Khan201729}. The unique permission and access-control management adopted in Corda is aligned with the Smart Money approach towards data sharing, particularly with the MTS scheme where payments become multilateral instead of bilateral transactions. One major difference between Corda and other permissioned DLT platforms is the adoption of Notary nodes. These are special purpose dedicated hosts which participate in the network and serve the purpose of time-stamping and preventing double-spends of funds.

Corda supports the use of accounts within a node, which act as logical partitions of the state objects which belong to that node. This enables hosting multiple entities within a single node and eliminating the need for each entity to host its own proprietary node. The Corda library accounts is used for two purposes: representing CBDC wallets and implementing permissioned data access control. With reference to Figure \ref{Fig.1}, CBDC wallets which act as bank accounts are assigned to Corda accounts. Thus, each wallet in a node is isolated from the other wallets in order to protect funds and personal data from eavesdropping or stealing. In the experiment, the \textit{SellerCWP} hosts multiple accounts for retail businesses and the BuyerCWP hosts accounts for consumers. For government departments, Corda accounts represent different administrative entities within an organisation, with a clear separation of duties. For HMRC in particular, there are two different accounts: VATpayments and VATInvestigator. The former is involved exclusively with validating payments with respect to the tax compliance element and the latter is used for requesting access to an account's data through a Smart Warrant.

An event, such as issuing new money, transferring money, or submitting a data access request represents an on-ledger event, or what is called a Corda state. This is the fundamental data block stored on the Corda ledger and constitutes an immutable record of a shared fact. When that fact has to be "updated", a new state is recorded, rendering the existing one as historic. The lifecycle of that fact over time is called a state sequence. In Smart Money, states represent paid and unpaid invoices, requested and accepted smart warrants and money issued to end users. Corda contracts are rulesets that define how a Corda state can be updated to a new one. A contract is the "control" which ensures that a transaction has been fulfilled according to the pre-defined conditions. If the conditions in the contract are fulfilled, the state update can take place and record itself on the ledger. In the opposite case, the state update is not permitted and the corresponding transaction fails. In MTS, contracts ensure that the VAT paid for an invoice matches the correct amount given the net value and VAT rate for the class of good or service sold, acting as a tax compliance actuator. Corda flows automate business processes by invoking a series of actions that cause state updates in a single execution.

When a supplier issues an invoice under MTS, their CWP initiates an invoice issue flow which executes all the necessary steps for completing the transaction in Corda such as creating a new state, verify signatures from interested parties, validate transaction against matching contracts and allocating funds to accounts. The Corda DL platform provides a toolset used to create a native token for representing assets that can be exchanged among the network participants. Smart Money employs the Corda Token SDK for representing financial support that is provided by government departments as part of benefit schemes, e.g. the Universal Credit scheme offered in the UK \cite{Omar2017}. These tokens are issued by government owned Corda nodes to Corda special purpose accounts which are hosted within CWP's (see sub-section \ref{issueprogmon}). The nodes are owned by government departments for which social benefits are within their remit e.g. the UK's Department for Work and Pensions. A beneficiary may hold both types of accounts: one for paying from a conventional wallet and one for paying from the token wallet. As this type of money is provided as a benefit scheme, it is subject to limitations with regards to the types of goods and services it can be spent on. An allowed goods list is prescribed for that purpose. A check against the allowed goods list is facilitated by the \textit{InvoiceContractTk} contract: whenever a wallet holder pays an invoice with tokens, the contract checks if the items in the invoice are part of the allowed goods list. If the check fails, the payment fails as well. This functionality is a demonstration of how Smart Money employs programmable money for supporting governance.

\subsection{MTS experiment participating entities}
A Smart Money experimental environment has been implemented for testing the feasibility of MTS. There are six participating nodes in the experimental network, each with a distinct role:
\begin{enumerate}
\item A Corda node which holds CBDC wallets for businesses that make sales labelled \textit{SellerCWP}.
\item A Corda node which holds CBDC wallets for citizens that make purchases labelled \textit{BuyerCWP}.
\item A Corda node which is owned by the HMRC labelled \textit{HMRCCWP}.
\item A Corda node which is owned by a LegalAuthority that authorises Smart Warrants labelled \textit{LegalCWP}.
\item A Corda Notary node for validating transactions labelled \textit{Notary}.
\end{enumerate}

For the purpose of this experiment, \textit{SellerCWP} contains only wallets for entities that conduct sales, whilst \textit{BuyerCWP} contains wallets for entities that conduct purchases. In reality, any CWP can host both types of wallets and facilitate any type of transactions for their customers. Moreover, each entity may participate with more than one node in the network to achieve high availability or boost performance under heavy workloads. The experimental scenario does not consider the communication exchange between the Smart Money system and the payments network; however, it employs a native digital currency (the CBDC) within Corda that is checked and updated for each transaction. 
When an invoice is issued or paid on the Corda ledger the CWP (seller's or buyer's) handles the task of passing on messages between the two layers.

\subsection{Issuing and paying an invoice}
\label{issuepayinv}
In MTS, payments are initiated by the Seller's CWP, who transmits a request-to-pay (R2P) message to the Buyer's CWP when the corresponding invoice is generated. That message includes the enhanced data element, such as an invoice in a structured data. Enhanced data is a capability which is already being provided by payment processors \cite{VISA2020} and is due to become standard in the UK in the years to come alongside R2P messaging \cite{Pay.UK}, using ISO20022. As soon as the Seller's CWP issues the R2P message, it also initiates a transaction in the Smart Money DLT.

That transaction involves the preparation steps on behalf of the seller node such as generating one-time keys, creating the \textit{SMInvoiceState}, signing the transaction and opening a session with each interested party. The interested parties, i.e. \textit{BuyerCWP} and HMRC then run the transaction against the \textit{InvoiceContract} and if the conditions in the contract are fulfilled, they proceed with signing the transaction and notify \textit{SellerCWP}. Then, the \textit{SellerCWP} shares the finalised \textit{SMInvoiceState} with HMRC and \textit{BuyerCWP} and submits the fully signed transaction to the \textit{Notary} node. The \textit{Notary} validates the timestamp of the transaction, notifies the interested parties to update their ledgers with the new DL state and returns a unique transaction id for the completed payment. When the transaction is completed, the ledger contains a new state which specifies an unpaid invoice which is only visible to the interested parties. 

\begin{figure}
    \includegraphics[width=\columnwidth]{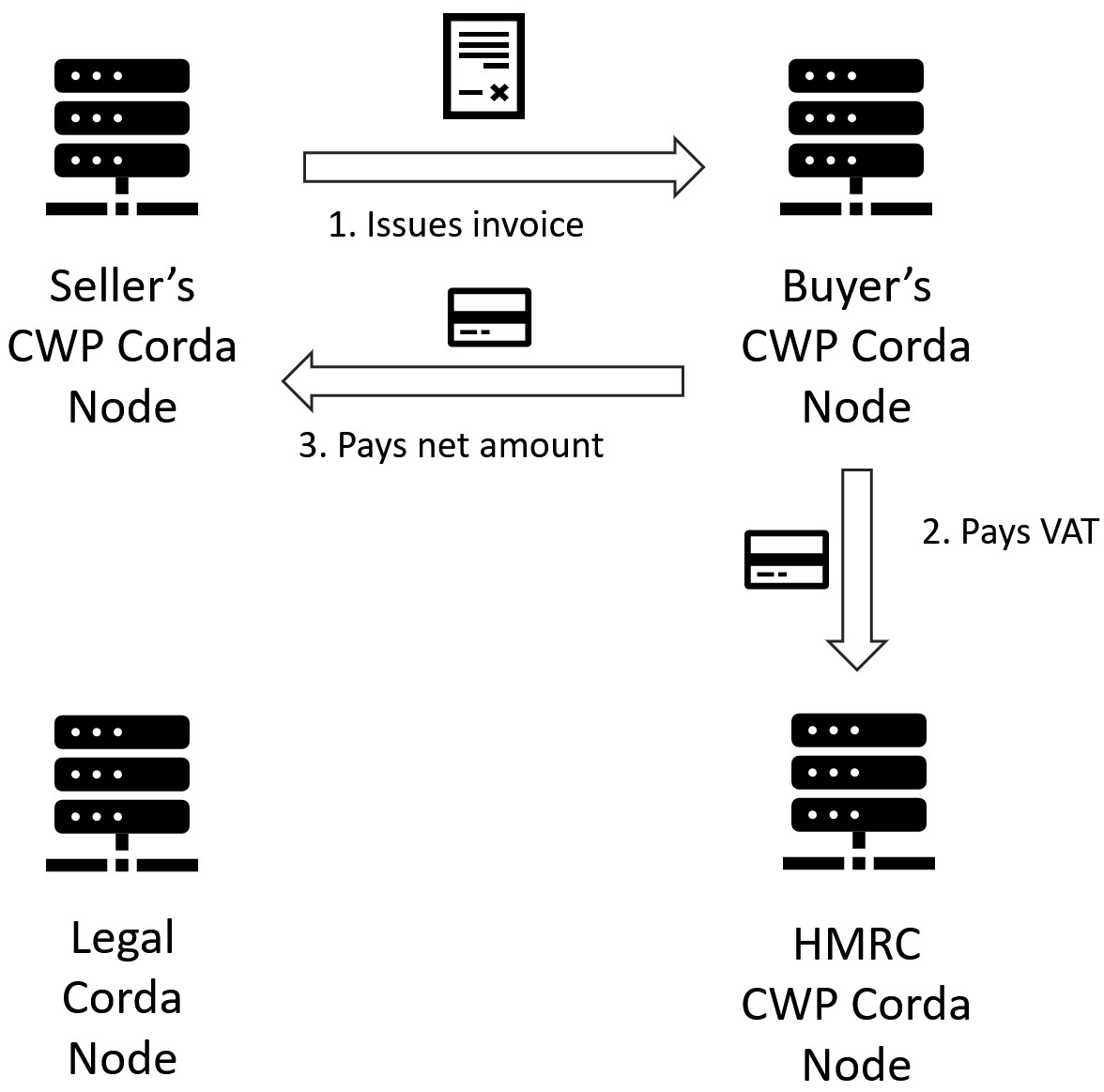}
    \caption{Issuing and paying an invoice}
    \label{Fig.3}
\end{figure}

To pay the issued invoice, the buyer responds to the R2P message through the Smart Money layer. The \textit{BuyerCWP} informs their customer of the pending payment and invoice through their CBDC Wallet User Interface, in the form of a smartphone or web banking app. The buyer has the chance to review the details of the payment and agree to the terms with the push of a button which grants permission to the \textit{BuyerCWP} to respond to the payment request. The \textit{BuyerCWP} executes the necessary preparation steps such as generating keys, creating the new \textit{SMInvoiceState} and signing the transaction. This is followed by calculating the amounts to be paid, i.e. the net amount to the seller's account in the \textit{SellerCWP} and the VAT amount to the HMRC's account in the \textit{HMRCCWP}. The flow then checks if the buyer's account has sufficient funds for fulfilling both parts of the payment and starts building the transaction for completing payment. At this stage, \textit{BuyerCWP} creates the new state, signs the transaction and initiates sessions with the interested parties who run the transaction against the InvoiceContract. If the conditions in the contract are fulfilled, they proceed with signing the transaction and notify \textit{BuyerCWP}. Then, the \textit{BuyerCWP} shares the finalised \textit{SMInvoiceState} with \textit{HMRCCWP} and \textit{SellerCWP} and submits the fully signed transaction to the \textit{Notary} node. The \textit{Notary} finalises the transaction by running the same steps as for issuing an invoice. The process is summarised in Figure \ref{Fig.3}.

\subsection{Requesting and Executing Smart Warrants}
\label{smartwarrants}
\begin{figure}
    \includegraphics[width=\columnwidth]{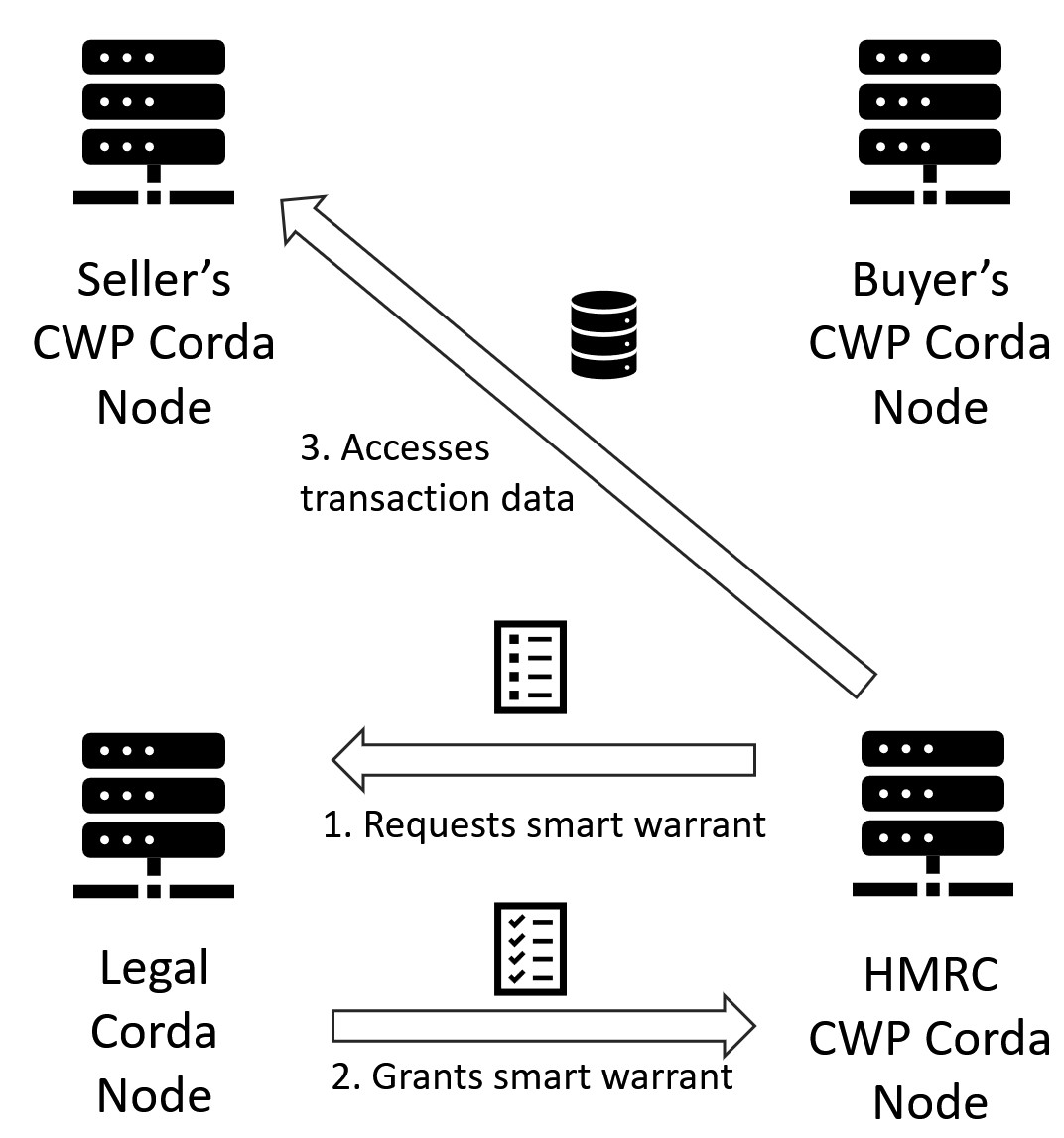}
    \caption{Requesting and granting Smart Warrant}
    \label{Fig.4}
\end{figure}

The purpose of Smart Warrants is to serve government departments for conducting fraud investigations and analytics upon a CBDC wallet holder's transasction history when reasonable grounds for criminal activity on behalf of the wallet holder exist. Conventionally, when similar access requests are conducted, invesigation is authorised by a legal authority, such as a magistrate. Smart Money incorporates a LegalAuthority node for that particular purpose: to review, authorise and audit data access requests to payments data. In the experiment, Smart Warrants are employed by a special purpose HMRC account named \textit{VATInvestigator}, for conducting investigations on the transactions of a CBDC wallet. The retrieved transactions are returned to the warrant requester account but they are not stored on the ledger in a new state. The Smart Warrant request transaction involves the preparation steps on behalf of the HMRC node such as generating one-time keys, creating the DataAccessRequest state, signing the transaction and opening a session with the LegalAuthority Node. LegalAuthority runs the transaction against the DataAccessContract. If the conditions in the contract are fulfilled, it signs the transaction and notifies \textit{HMRCCWP}. The \textit{VATInvestigator} account also signs the transaction, submits it to the \textit{Notary} node for validation. At the last stage, the \textit{Notary} notifies the parties involved to update their ledger copies and returns a transaction ID for the warrant request. The warrant requester can execute the warrant when they deem necessary. The Smart Warrant must be executed by the requester account, otherwise the HMRC node rejects the execution. The requester account must provide the ID and LegalAuthority that issued the warrant. If these fields don't match with an existing issued warrant, then the HMRC node rejects the execution. If the execution passes this test, then a new DataAccessRequest state is recorded which uses the existing DataAccessRequest state as input. This new state, which contains a timestamp of execution is appended to a transaction which is signed by HMRC and passed on to LegalAuthority node for validation and signing. The \textit{Notary} node is notified, retrieves the data from the investigated account's CWP ledger, returns the data on the requester's user interface and updates the ledgers of HMRC and LegalAuthority with the new confirmed state. Smart Warrants provide SupTech capabilities for governments, to ensure that investigations on personal data are performed lawfully and transparently. The process is summarised in Figure \ref{Fig.4}.

\subsection{Issuing programmable CBDC to wallets}
\label{issueprogmon}
\begin{figure}
    \includegraphics[width=\columnwidth]{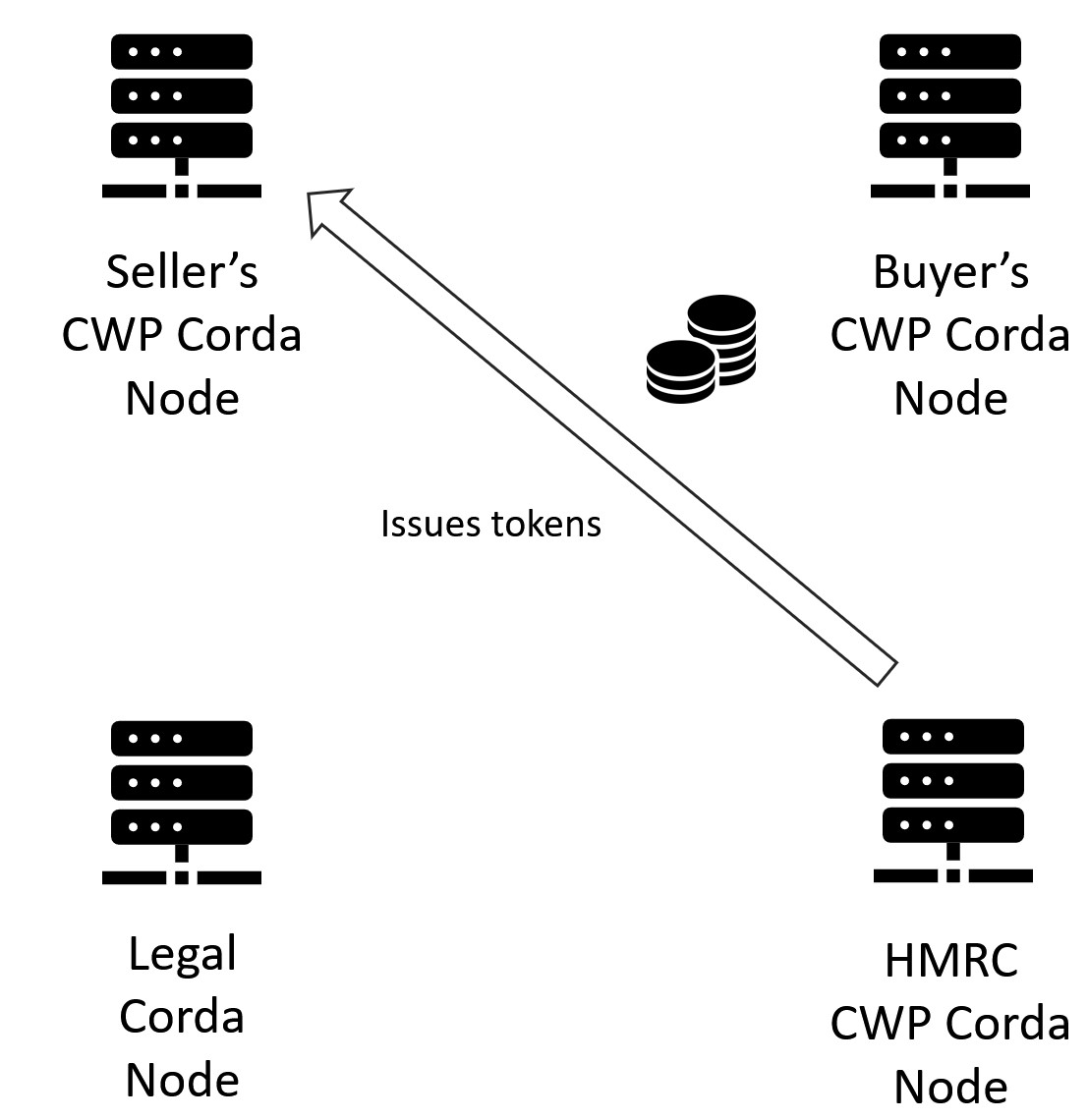}
    \caption{Issuing programmable money}
    \label{Fig.5}
\end{figure}

In addition to operating as an information management layer to the payment systems, the Smart Money DLT can facilitate payments with the use of a native CBDC which resides on the ledger. In this case, the exchange of funds takes place exclusively outside the payments network, however the VAT for transactions is still paid automatically through MTS. For the purpose of the experiment, the Token SDK has been incorpotrated in Smart Money to implement financial support along the lines of the Universal Credit scheme \cite{Royston2012} which is operating in the UK. This scheme, which is handled by the Department of Work and Pensions, is operated under the HMRC for the purpose of the experiment and it is controlled through the \textit{HMRCCWP} node. The tokens issued by the HMRC can be used in transactions to pay for goods within an allowed goods list. To issue tokens to a wallet, the HMRC node initiates an Issue Invoice flow which executes the necessary steps. This flow can only executed by the \textit{HMRCCWP} hence no other node can issue these tokens within the Smart Money DLT network. The flow assigns new tokens to the specified wallets sequentially, through a session that is initiated between \textit{HMRCCWP} and \textit{BuyerCWP}. Effectively, HMRC instructs \textit{BuyerCWP} to update the balance on the user's Corda account i.e. the receiver's CBDC wallet. Similarly, this process can be executed between \textit{HMRCCWP} and any CBDC wallet provider. The process is summarised in Figure \ref{Fig.5}.

\section{Evaluation methodology}
\subsection{Performance factors and indicators for DLT feasibility and scalability studies}
DLT-based applications for a multitude of domains are currently being researched, with a focus on demonstrating the feasibility and scalability of these solutions. DLTs which are based on either permissoned or permissionless access models are assessed with respect to how certain application requirements affect several performance metrics of the system. The majority of feasibility studies on the application of DLTs are concerned with how several operating factors affect a number of performance metrics.

A literature review has been conducted which shows that the majority of studies emphasise on throughput and latency as there is a common trade-off between data integrity and processing speed for DLT solutions. Of particular interest is the capability of a DLT application to scale with increasing volumes of workload. Workload is assessed in terms of how many transactions are submitted to the ledger per unit time \cite{Jiang2020, Huang2020, Melo2019, Khan2020, Sharma2018, Novo2018, Hao2018, Kuzlu2019, Wang2019b, Pal2020}  as well as the size of the data which is processed for each transaction \cite{Khan2020, Sharma2018, Novo2018, Wang2019b, Si2019, Qinghua2019, Chen2019, Li2019}. Other works study the effect of the employed consensus mechanism (Proof-of-Work, Proof-of-Stake, Byzantine Fault Tolerance) \cite{Khan2020, Sharma2018, Novo2018, Hao2018, Kuzlu2019}. The number of participating entities can also have an effect in performance \cite{Khan2020, Sharma2018, Novo2018, Kuzlu2019, Wang2019b, Si2019, Qinghua2019, Li2019}, as more nodes and users present a delimiting effect in transaction throughput \cite{Jiang2020, Huang2020, Melo2019, Khan2020, Sharma2018, Novo2018, Hao2018, Kuzlu2019, Si2019, Chen2019, Huang2019}
and latency \cite{Jiang2020, Huang2020, Melo2019, Khan2020, Sharma2018, Novo2018, Hao2018, Wang2019b, Pal2020, Si2019, Qinghua2019}. It has also been found that when the number of transactions becomes very high, some of those transactions might fail, it is therefore useful to measure the loss rate as a performance metric \cite{Jiang2020, Huang2020, Melo2019, Khan2020}. Table \ref{factorsandmetricsinlit} summarises the evaluation metrics for several DLT feasibility and scalability studies found in scientific literature.

The study presented in this paper, investigates the impact of the following factors on scalability:  Transaction Load (\textit{tx}), Data Volume (\textit{vol}), Number of clients (\textit{cl}). Transaction Load is defined as the number of invoices that have been submitted for issue or payment in a run. Data Volume refers to the number of items that are exchanged during the transaction and are included in the corresponding invoice. We refer to clients as the number of endpoints that communicate with one of the nodes participating in the Smart Money system and submit transactions to be issued or paid. Within the context of a CBDC, the term "endpoints" may refer to: a) end-user devices which hold an account holder's private key (plastic card, mobile phone) and interact directly with the CWP's Corda node or b) a special-purpose "aggergator" server which is operated by a CWP and collects transaction requests from end-users and submits them to the Corda node in bulk. By varying each of the factors (which constitute the independent variables of the experiment), the effect on scalability can be measured and analysed. Regarding the effect of the consensus mechanism on scalability, Corda does not use a consensus system. Instead, it provides a reliable witness, by means of \textit{Notary} nodes which prevent double-spending \cite{R32023}. Therefore, the impact of the consensus mechanism is not considered in the experiment.

With reference to the performance metrics, results for the variable Transaction Rejection Rate are not presented, as Corda has consistently completed all the transactions throughout the tests conducted. Therefore, only Throughput (\textit{tr}) and Latency (\textit{la}) have been selected as the measurement variables for the tests. The performance factors and metrics are summarised in table \ref{factorsandmetrics}.

\begin{table*}

\begin{tabular*}{\textwidth}{@{}l@{\extracolsep{\fill}}cccccccc}
\hline
& & & Factor & & & Performance metric & \\
\hline
Source & Transaction & Data & Consensus & No of clients & Throughput & Latency & Transaction  \\
& Load & volume & mechanism (*) & & & & rejection rate (*) \\
\hline
\cite{Jiang2020, Huang2020, Melo2019} & X & & & & X & X & X \\
\cite{Khan2020, Sharma2018, Novo2018} & X & X & X & X & X & X & \\
\cite{Hao2018} & X & & X & & X & & \\
\cite{Kuzlu2019} & X & & X & X & X & &  \\
\cite{Wang2019b} & X & X & & X & & X \\
\cite{Pal2020} & X & & & & & X & \\
\cite{Si2019} & & X & & X & X & X & \\
\cite{Qinghua2019} & & X & & X & & X & \\
\cite{Chen2019} & & X & & & X & & \\
\cite{Li2019} & & X & & & & X & \\
\cite{Huang2019} & & & & X & X & X & \\
\hline
\multicolumn{8}{l}{(*) denotes factors and metrics not considered in this study} \\
\end{tabular*}
\caption{Factors and Metrics in the literature}
\label{factorsandmetricsinlit}
\end{table*}

\subsection{Experimental design}
The purpose of the experiment is to demonstrate that the Smart Money Corda DLT is a feasible and scalable design for automated VAT split payments with a Central Bank Digital Currency. We examine the efficiency of smart contracts to fulfil VAT split payments and programmable money to prove the feasibility of the design. We also measure the effects caused to system performance by applying variable levels of stress (through the various factor levels) in order to demonstrate the design’s scalability. To assess the effect from each individual factor, tests are conducted with varying values for one factor while the remaining two factors are held constant. This approach allows for evaluating each of the factors separately and understanding how, and to what extent they affect performance. This is important for informing future iterations of the Smart Money CBDC as well as other, third-party CBDC implementations which are based on R3 Corda and adopt tax split-payments.

To investigate the impact of automatic VAT payments, the experiment, the tests will be conducted twice: once with split payments enabled and once with split payments disabled. We consider scalability as the ability to support increasing workloads with reference to the chosen performance factors. Scalability is demonstrated when performance decreases linearly with workload as in \cite{Manzoor2021, Luu2016254, Hafid2020, Zhou2020}. Scalability tests involved with the impact of Universal Credit scheme have not been conducted. This is because the check for disallowed items in an invoice is made irrespective of the money instrument (account balance or Universal Credit tokens) used for the corresponding payment. Therefore, paying through Universal Credit has no impact on scalability.

\subsection{Smart Money Corda DLT environment}
\label{SMCordaDLTEnv}
The feasibility and scalability of the Smart Money CBDC is evaluated in a controlled experimental environment which simulates a real-world scenario that involves payments occurring between counterparties as well as the information exchange among all the interested stakeholders. The tests have been conducted using Corda Open Source v4.3 as this version was available at the time of development and subsequent versions do not provide improvements which impact on the objectives of the experiment. The simulation environment consists of a computer network where each of the Making Tax Smart use case entities, participates by means of a Corda node. There are seven Corda nodes in total, which are executed as Java Virtual Machines from within an equal number of Windows 2012 virtual machines. The hypervisor is VMWare vSphere 6.0. The setup is running on a 16-core Intel Xeon E5-2460 v3 @ 2.6GHz and a total of 256GB of RAM which is allocated to the hyperivsor and VM’s. Three of the nodes (HMRCCWP, SellerCWP, BuyerCWP) are allocated more hardware resources (by means of CPU cores and RAM) as preliminary tests demonstrated that this produced performance gains, as these nodes execute the most resource-consuming flows. Moreover, the machines which host \textit{BuyerCWP} and \textit{SellerCWP} also host the RPC clients which generate payments (issue and pay) thus further increasing the resource requirements for those two machines. The remaining nodes, which execute less-demanding tasks are allocated less hardware resources. Table \ref{cordanodes} summarises the list of participating nodes, their description and their hardware specifications within the Corda experimental network.

The full functionality of MTS has been implemented in a CorDapp. Each participant node contains the part of the CorDapp which is relevant to their purpose. Similarly, \textit{BuyerCWP} and \textit{SellerCWP} do not execute the code required for requesting smart warrants from node \textit{LegalCWP}. The functionality of each node is presented in Table \ref{cordanodes}. Corda nodes are not configured to communicate when they are built from source, i.e. the IP addresses of the remaining nodes are not included in their configuration files. Corda’s native network bootstrapper is used for updating the node configuration with the list of IP’s of subsequent nodes. After that process is completed, the nodes are ready to be started up as JVM’s. Users map to Corda accounts therefore any request submitted to a node (e.g. for issuing a payment, for paying an invoice, for executing a smart warrant etc) is done on behalf of an account. Accounts from different nodes interact with each other only through the nodes they are registered on. An account’s parent node is responsible for transmitting the message to the destination node which in turn, notifies the destination account. An RPC (Remote Procedure Call) client application is used for interfacing a user/account with a node. That RPC is submitted to the target node through the network hence the user can reside in any network station that can communicate with the node through TCP/IP. For the purpose of this experiment, the users reside within the node they are registered on i.e. requests for payment are submitted from clients executed within the same VM which hosts the \textit{SellerCWP}. Similarly, payments are submitted from clients executed within the same VM which hosts the \textit{BuyerCWP}.

\begin{table}[ht!]
\begin{tabu} to \columnwidth {p{2cm}|p{6cm}}
\hline
\textbf{Node name} & \textbf{Functionality} \\
\hline
HMRCCWP & Validates payments for invoices issued from accounts registered to SellerCWP and paid by accounts registered to BuyerCWP. \\
& Issues UC money (tokens) to accounts registered to BuyerCWP. \\
& Requests smart warrants from LegalCWP for accounts registered to SellerCWP. \\
& Executes smart warrants authorised by LegalCWP. \\
& Creates accounts to BuyerCWP, SellerCWP, LegalCWP and shares those accounts with other nodes. \\
& Hosts VATPayments account (for payment validation). \\
& Hosts VATInvestigator account (for smart warrant execution) \\
\hline
BuyerCWP & Pays invoices on behalf of accounts registered to BuyerCWP which are issued by accounts registered to SellerCWP. \\
& Hosts buyer accounts which hold balances (in Current and UC accounts). \\
\hline 
SellerCWP & Issues invoices on behalf of accounts registered to SellerCWP which are to be paid by accounts registered to BuyerCWP. \\
& Hosts seller accounts which hold balances (in Current accounts). \\
\hline
LegalCWP & Authorises smart warrants issued from the account registered to HMRCCWP. \\
& Hosts LegalAuthority account (for smart warrants authorisation). \\
\hline
Notary & Ensures uniqueness to transactions through validation. \\
\hline
\end{tabu}
\caption{Corda nodes}
\label{cordanodes}
\end{table}

\begin{table*}[!ht]
\setlength{\tabcolsep}{1.5pc}
\settowidth{\digitwidth}{\rm 0}
\catcode`?=\active \def?{\kern\digitwidth}
\begin{tabular*}{\textwidth}{p{1.5cm}|p{7cm}|p{1cm}|p{0.5cm}|p{1cm}}
\hline
\hline
Node name & Description & CPU cores & RAM size & IP address \\
\hline
HMRCCWP & Operated by HMRC for validating payments and conducting investigations through smart warrants authorised by LegalCWP. & 8 & 16GB & 192.168.1.110 \\
\hline
SellerCWP & Operated by a CWP which hosts accounts of businesses which make sales. & 8 & 64GB & 192.168.1.111 \\
\hline
BuyerCWP & Operated by a CWP which hosts accounts of consumers which make purchases. & 8 & 64GB & 192.168.1.112 \\
\hline
LegalCWP & Operated by a Legal authority for authorising smart warrants to HMRC & 4 & 4GB & 192.168.1.113 \\
\hline
Notary & Essential Corda node which validates transactions before they are added to the ledger. & 4 & 4GB & 192.168.1.114 \\
\hline
\end{tabular*}
\caption{Corda nodes in the experimental setup}
\label{expsetup}
\end{table*}

\begin{table*}
\begin{tabu} to \textwidth {p{4.5cm}|p{12cm}}
\hline
\textbf{Factor} & Description \\ 
\hline
Transaction load \textit{tx} & The number of transactions submitted to the ledger.  \\
\hline
Data volume \textit{vol} & The number of items on an invoice. \\
\hline
No of clients \textit{cl} & The number of clients that interact with the ledger via a correspondent node. \\
\hline
\hline
\textbf{Metric} & Description \\ 
\hline
Throughput \textit{tr} & The number of successful transactions completed per unit time. Measured in Transactions per Second (TPS). \\
\hline
Latency \textit{la} & The time delay between two events: 1) transaction issued and 2) transaction validated. Measured in seconds. \\
\hline
\end{tabu}
\caption{Factors and Metrics}
\label{factorsandmetrics}
\end{table*}

\section{Feasibility study}
The feasibility of Smart Money system was evaluated using real-life scenarios where the transactions involve buyers who purchase goods from sellers and where HMRC utilises the Smart Money system for enforcing tax collection and compliance. There are seven classes of goods sold in these transactions where each class incurs a VAT rate i.e. Adult Clothing:20\%, Alcohol: 20\%, Books: 0\%, Children’s Clothing: 0\%, Electrical: 20\%, Energy: 5\%, Groceries: 0\%. A transaction involves the following entities:
\begin{enumerate}
\item A seller named MegaCompany, which is a retailer that sells all the available classes of goods.This seller takes part in MTS through their account which is held in a Corda node named \textit{SellerCWP}. This node hosts Corda accounts for sellers.
\item A buyer named Alice, which is a consumer that may purchase any of the available classes of goods. This buyer takes part in MTS through their account which is held in a Corda node named \textit{BuyerCWP}. This node hosts Corda accounts for buyers.
\item HMRC, which collects VAT for all VAT-incurring transactions that take place through MTS. HMRC takes part in MTS through an account named VATPayments which is held in a Corda node named HMRCCWP.
\item The Notary node which, by design, participates in every transaction which takes place in R3 Corda, for validation purposes.
\end{enumerate}

Each core functionality of Smart Money is implemented using Corda flows (see Section 4). The feasibility is evaluated in the form of assertion tests \cite{Rosenblum1995} which check the correctness of the flows i.e. whether the flows produce the correct results. The assertion tests presented in the remainder of this section verify that it is feasible to implement the MTS core functionality with Distributed Ledger Technology. 

\subsection{Issuing and paying an invoice}
This Section covers the scenarios for testing the feasibility of issuing and paying invoices with VAT split payments within MTS. These scenarios cover the cases where an invoice is being paid by an account which has or has not sufficient funds to cover the total value of the goods purchased.

\begin{figure}
\includegraphics[width=\columnwidth]{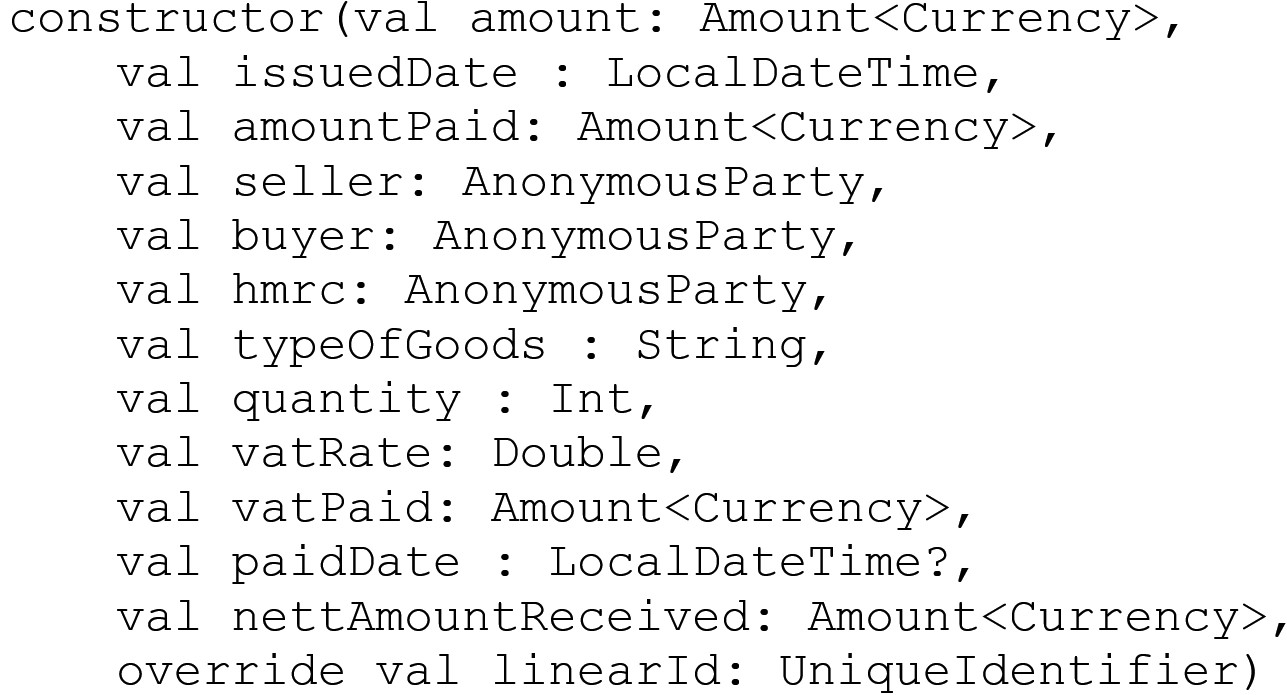}
	\caption{\label{Fig.2} Invoice state constructor}
\end{figure}

In this test (see Algorithm \ref{alg:paysuff}), a payment is created when a seller (MegaCompany) issues an invoice to a buyer (Alice) (See row 8 in Algorithm \ref{alg:paysuff}. This invoice contains information which is found at a regular invoice such as the class of goods purchased, the quantity of items per class, the net price, the VAT rate for each class of goods and the total price for the invoice. In addition, each invoice contains the type of money that will be used for the payment which can be one of two types: Current and Token. If the type is Current, the invoice can only be paid with funds from Alice’s ’regular’ account. That’s because certain item classes (e.g. Alcohol) do not belong in the allowed goods (see Section \ref{sec:sample:appendix} in the Appendix). For a complete list of fields, see the invoice state constructor in Figure \ref{Fig.2}.
After the invoice is issued, it is appended in the unpaid invoice queue for both Alice and MegaCompany. HMRC, via its \textit{VATPayments} account should not have visibility of any unpaid invoices, hence is not aware of the transaction between Alice and MegaCompany. The first assertion test (AT1) verifies that the (yet unpaid) transaction is only visible by the accounts Alice and MegaCompany (see rows 9-12 in Algorithm \ref{alg:paysuff}). The invoice is paid by Alice which is followed by the next assertion test (AT2) which verifies that all counterparties see the invoice as paid (rows 14-17 in Algorithm \ref{alg:paysuff}. The final assertion test (AT3), verify that all account balances are now updated according to the transactions (payment for goods and VAT split payments, see rows 18-21 in Algorithm \ref{alg:paysuff}). A subsequent test (see Algorithm \ref{alg:payinsuff}) was conducted to verify that a transaction cannot be completed if Alice does not have sufficient funds in their account to pay for the invoice’s total amount. In this case, the unpaid invoice is created first and the payment fails with an error "insufficient funds available" when Alice tries to pay for the invoice. The invoice, according to Assertion Test 4 (AT4) must remain in the unpaid state (see rows 10-13 in Algorithm \ref{alg:payinsuff}). The account balances for all counterparties should also remain unchanged (see rows 14-17 in Algorithm \ref{alg:payinsuff}), which is verified by the Assertion Test 5 (AT5).

The next series of assertion tests (as listed in Table \ref{assertiontestst}) test was conducted to verify that goods can be purchased by spending tokens. In this test, a different shopping list was used (see listing \ref{listing2}), where only goods from the allowed goods list are to be purchased. According to the Assertion Test (AT6), the invoice state must change to paid (see Algorithm \ref{alg:paywithtokens} rows 10-13). Assertion Test 7 (AT7) was conducted to check whether the balances are correct after the tokens were spent (see Algorithm \ref{alg:paywithtokens} rows 14-17). One more test (see Algorithm \ref{alg:paywithtokensnotallowed}) was conducted to verify that a transaction cannot be completed if Alice pays for goods that are not in the allowed goods list e.g. Alcohol, with money from their token account (see listing \ref{listing2} for the shopping list). In this case, the payment for the invoice fails with an error "you cannot pay for invalid goods with money from your token account". The invoice remains in the unpaid state (see rows 10-13 in Algorithm \ref{alg:paywithtokensnotallowed}). The account balances for all counterparties should also remain unchanged (see rows 14-17 in Algorithm \ref{alg:paywithtokensnotallowed}). The assertion tests returned positive results for all the test cases thereby proving the split VAT payment functionality is feasible with the MTS use case for the Smart Money system. This has been found to work properly for variable VAT rates, in the same invoice, e.g. an invoice with goods of all three VAT rates (0\%, 5\% and 20\%).

\subsection{Requesting and issuing warrants}
This Section covers the scenarios for testing the feasibility of establishing controlled access to payments data for government agencies, after authorisation by a legal authority. These scenarios cover the case where HMRC conducts an investigation to a retailer and requests access to payments that have been completed through MTS. In this test case (see Algorithm \ref{alg:smartwar}), a smart warrant is created when an investigator who works for HMRC (\textit{VATInvestigator}) submits a data access request to a legal entity (LegalAuthority) so that they gain access to the payments data of a seller account (MegaCompany). The LegalAuthority signs the request (this process is done automatically in the feasibility tests) and returns a signed Data Access Request (DAR) which can only be executed by the \textit{VATInvestigator} account and can only be used for accessing the payments data of MegaCompany’s account. The first assertion test (AT10) verifies that the signed DAR has been created and is a validated, unexecuted request authorised by the LegalAuthority (see rows 5-7 in Algorithm \ref{alg:smartwar}). When the \textit{VATInvestigator} executes the warrant, the warrant’s state should first become executed (for the second assertion test, AT11, see rows 9-10 in Algorithm \ref{alg:smartwar}). This account can then execute the warrant, which runs a vault query on the MegaCompany’s database of transactions and fetches all results on the \textit{VATInvestigator}’s screen (see row 12 in Algorithm \ref{alg:smartwar}). This concludes the third assertion test (AT12). The final assertion test for Smart Warrants (AT13), verifies that the \textit{VATInvestigator} account can only execute this query once (row 13 in Algorithm \ref{alg:smartwar}). The feasibility tests confirm that the functionality required for a CBDC with MTS capabilities is feasible using the Smart Money system, particularly implemented with the R3 Corda DLT.

\begin{table}
\begin{tabu} to \columnwidth {p{1cm}|p{5.5cm}|p{1cm}}
\hline
\textbf{Test ID} & \textbf{Description} & \textbf{Test result} \\ 
\hline
& Unpaid invoice is stored in MegaCompany ledger & True \\
AT1 & Unpaid invoice is stored in Alice's ledger & True \\
& Unpaid invoice is not stored in VATPayment's ledger & True \\
\hline
& Invoice state in MegaCompany's ledger is Paid & True \\
AT2 & Invoice state in Alice's ledger is Paid  & True \\
& Invoice state in VATPayment's ledger is Paid  & True \\
\hline
& MegaCompany's account balance is credited with Netamount & True \\
AT3 & Alice's account balance is debited with Totalamount & True \\
& VATPayment's account balance is credited with VAT & True \\
\hline
& Invoice state in MegaCompany's ledger is not Paid & True \\
AT4 & Invoice state in Alice's ledger is not Paid  & True \\
& Invoice state in VATPayment's ledger is not Paid  & True \\
\hline
& MegaCompany's account balance remains unchanged & True \\
AT5 & Alice's account balance remains unchanged & True \\
& VATPayment's account balance remains unchanged & True \\
\hline
& Invoice state in MegaCompany's ledger is Paid & True \\
AT6 & Invoice state in Alice's ledger is Paid  & True \\
& Invoice state in VATPayment's ledger is Paid  & True \\
\hline
& MegaCompany's account balance is credited with Netamount & True \\
AT7 & Alice's account balance is debited with Totalamount & True \\
& VATPayment's account balance is credited with VAT & True \\
\hline
& Invoice state in MegaCompany's ledger is not Paid & True \\
AT8 & Invoice state in Alice's ledger is not Paid  & True \\
& Invoice state in VATPayment's ledger is not Paid  & True \\
\hline
& MegaCompany's account balance remains unchanged & True \\
AT9 & Alice's account balance remains unchanged & True \\
& VATPayment's account balance remains unchanged & True \\
\hline
AT10 & Signed DAR is created and unexecuted & True \\
\hline
AT11 & Signed DAR is executed & True \\
\hline
AT12 & Fetched data is MegaCompany's actual transactions & True \\
\hline
AT13 & VATInvestigator can only query MegaCompany's transactions once & True \\
\hline
\end{tabu}
\caption{Assertion tests}
\label{assertiontestst}
\end{table}

\subsection{Findings of feasibility study}
The results presented in this Section have confirmed that designing and developing a CBDC with VAT split payment capabilities is feasible using Distributed Ledger Technology and smart contracts, particularly the R3 Corda open-source blockchain platform. It has been also shown that the Smart Money DLT provides the required capabilities for investigations on behalf of HMRC, by way of smart warrants, with the necessary safeguards in place to assure citizens' rights to privacy. The Smart Money System can even act as an enabler of smart policies through a CBDC, where social services such as the UK's Universal Credit scheme can be offered more efficiently and transparently, which makes it more beneficial for citizens, government as well as for businesses.

\section{Scalability study}
This Section presents the results of the scalability study for the Smart Money CBDC which tests the effects of three factors to system performance. Particularly, it has to be shown that the performance decrease is tractable, as the factors shift from lower values to higher values. There is a question about what is the upper bound for these factors within the context of this scalability study. The selection of the value ranges for each of those factors in the experiment takes into account hardware limitations of the employed testbed and real-life use cases.

Given that a single RPC client may utilise significant portions of RAM (between 128MB and 1GB), experimental runs employing more than 100 clients become infeasible, since the VM's RAM resources will be exhausted (see subsection \ref{SMCordaDLTEnv} for the use of RPC clients as transaction generators). It therefore applies that $10\leq cl\leq 100$.

 In accordance with the basket of goods and services study published by the Office for National Statistics \cite{Gooding2022}, the value 100 is used as the maximum number for the Items per invoice performance factor, hence $10\leq vol\leq 100$.

 Transaction load \textit{tx}, i.e. number of invoices issued and paid is a factor which depends on the number of clients \textit{cl}, because the total number of invoices is equal to the number of clients multiplied by the number of invoices (issued or paid) per client. The effective upper bound for Transaction load in the experiments was set to 10,000 invoices (e.g. 100 clients * 100 invoices per client and 10 clients * 1000 invoices per client) for all client transactions. Preliminary tests with higher transactions loads were conducted and the results were consistent with those illustrated in subsequent sections. These results are omitted for the sake of brevity. Every data point in the results presented in Figures \ref{Fig.6}, \ref{Fig.7} and \ref{Fig.8} represents an average of 10 runs for each test case.

\subsection{Impact of data volume (\textit{vol}) and number of clients (\textit{cl})}A series of experiments was conducted to assess the impact of increasing data volume on the Smart Money DLT. A three-level scale of data volume with reference to the number of items included within an invoice was adopted, i.e. 10 items per invoice, 50 items per invoice and 100 items per invoice. The total number of invoices was fixed to the value 1000. The experiments were repeated for various number of clients [10, 20,...,100] and the transaction load i.e. the total number of invoices was spread among the clients so that the following formula is satisfied:$cl$ x $tx$ per client $= 1000$.

This design makes it possible to evaluate the impact of increasing clients in the network in addition to the effects of increasing data volume. Furthermore, all tests are executed twice; one set of runs with split-payments enabled and one set of runs without split payments, where the full amount is paid directly to the seller with no involvement from the HMRC.

\begin{figure*}
\includegraphics[width=\textwidth]{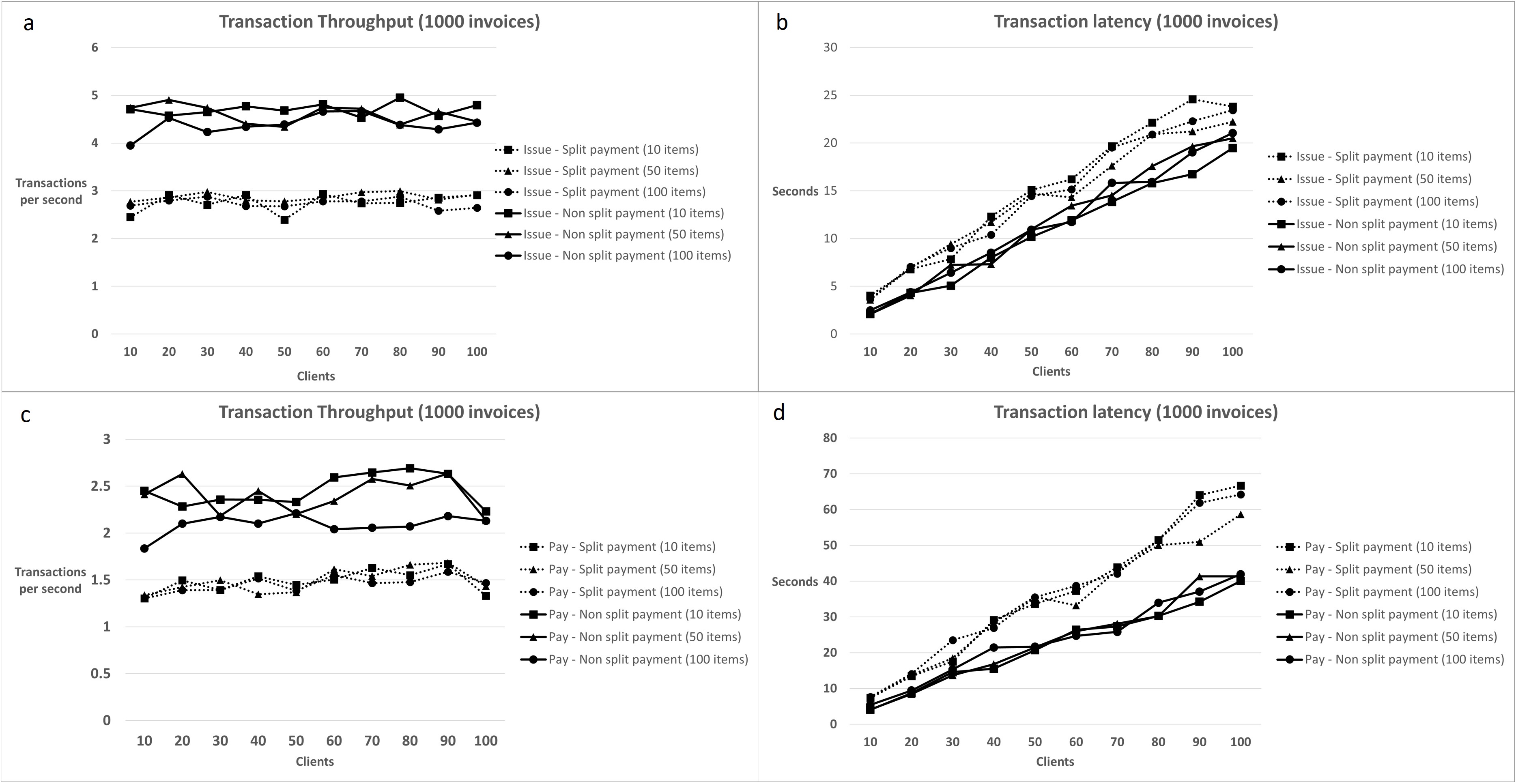}
	\caption{\label{Fig.6} a. Transaction throughput - issue [1000 invoices], b. Transaction latency - issue [1000 invoices], c. Transaction throughput - payment[1000 invoices], d. Transaction latency - payment [1000 invoices]}
\end{figure*}

The results in Figure \ref{Fig.6} show that the Smart Money DLT achieves similar performance for all data volume levels. By inspecting the families of curves (Issue-Split, Issue-Non split, Pay-Split, Pay-Non split) there is no clear within-family separation, which shows that the number of items per invoice has no effect on performance when invoices are issued or paid. Additional findings can be made with regards to the number of clients in transactions. In Figure \ref{Fig.6}a the curves follow almost a straight line which shows that throughput is not affected by increasing the number of clients. In the case of payments (Figure \ref{Fig.6}c), throughput appears to slightly increase when the transaction load is spread among more clients, which demonstrates high scalability. Transaction latency (\ref{Fig.6}b, \ref{Fig.6}d), increases with the number of clients, albeit linearly. Overall, the results show that the Smart Money DLT demonstrates scalability when the data volume of transactions increases. Similar results are observed when increasing the number of clients in the network. The results also show that with split-payments enabled, performance (in terms of transaction throughput and latency) drops by a factor of 1.5 to 2. This is caused by the extra computational burden required for calculating and transferring the VAT of a payment to HMRC for either issuing or paying an invoice.

\subsection{Impact of transaction load (\textit{tx})}
There are two ways for varying \textit{tx}, i.e. a) increase the amount of invoices per client while keeping the client population constant and b) increase the number of clients while keeping the amount of invoices per client constant. A separate series of tests was conducted for each of the two approaches. The results from subsection 7.1 confirm that the data volume has no impact on performance therefore it was deemed unnecessary to control this factor for the remainder of the experiments. The number of items per invoice was set to 10 and is held constant across all tests. 
\subsubsection{Varying invoices per client and keeping clients \textit{cl} constant}
\begin{figure*}
\includegraphics[width=\textwidth]{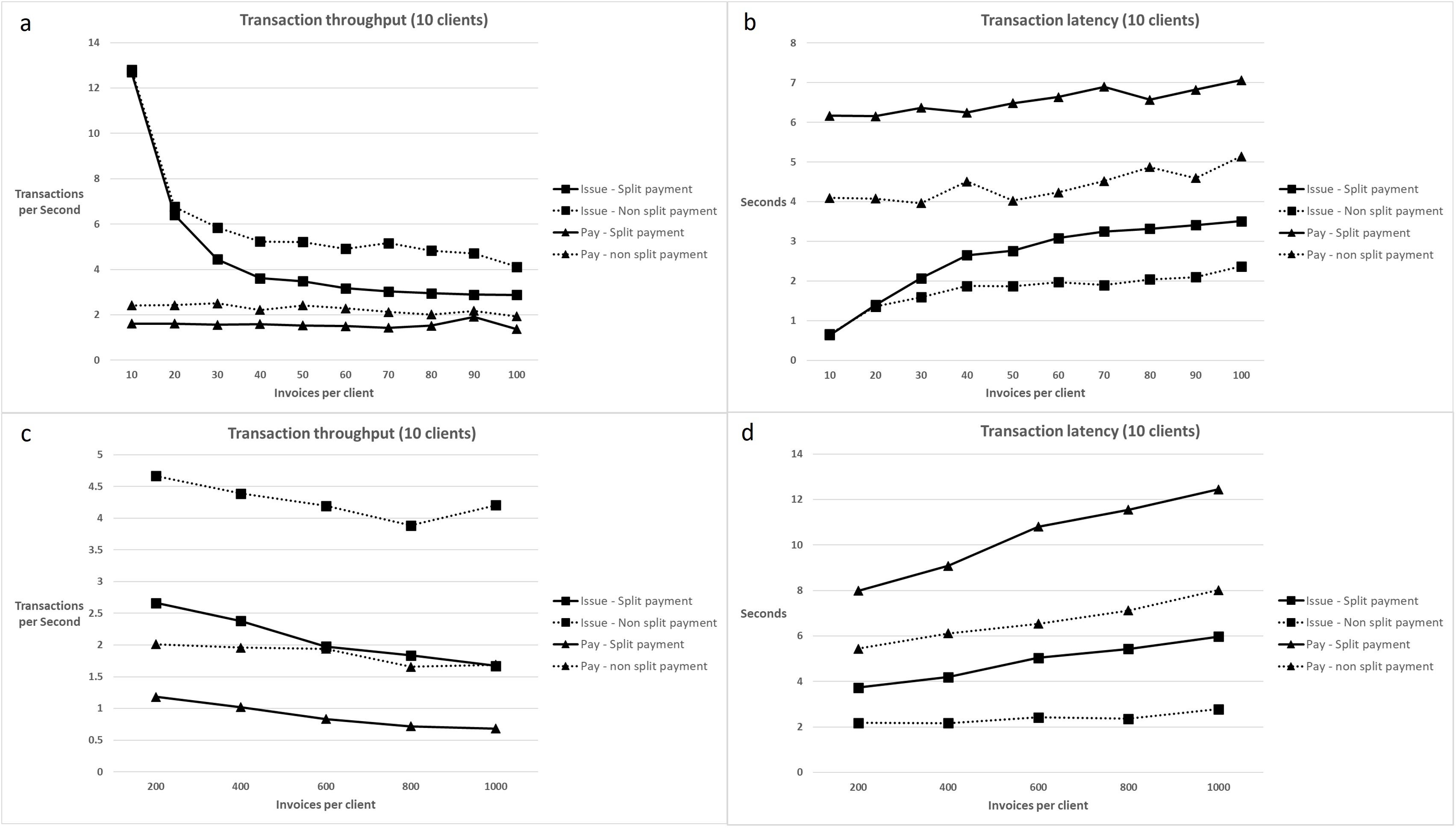}
	\caption{\label{Fig.7} a. Transaction throughput [10 clients], b. Transaction latency [10 clients], c. Transaction throughput [10 clients], d. Transaction latency [10 clients]}
\end{figure*}

In this set of experiments there were 10 clients issuing and paying invoices at the same time. The results illustrated in Figure \ref{Fig.7}a and Figure \ref{Fig.7}b show that performance drops instantly from 10 to 20 invoices per client. From there, performance (in terms of throughput and latency) drops smoothly as the number of invoices increases. This shows the Smart Money DLT (i.e. the MTS) scales appropriately with increasing the number of payments (issued and paid) by holding the number of clients constant and increasing the number of transactions per client. This is also observed when the number of invoices increases further, from 200 to 1000 per client, see Figure \ref{Fig.7}c and Figure \ref{Fig.7}d. Enabling VAT split payments impacts on performance in this case as well, since transaction throughput with split payments is always lower with an associated higher latency.

\subsubsection{Varying number of clients (\textit{cl}) and keeping invoices per client constant}
\begin{figure*}
\includegraphics[width=\textwidth]{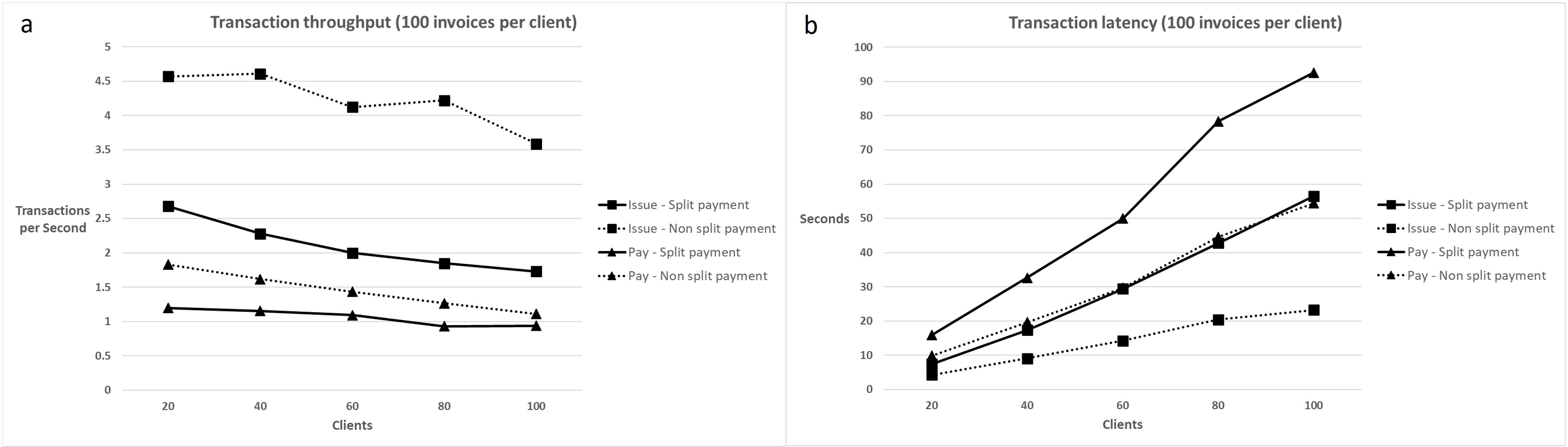}
	\caption{\label{Fig.8} a. Transaction throughput [100 invoices per client], b. Transaction latency [100 invoices per client]}
\end{figure*}

In this set of experiments, each client was configured to submit 100 invoices both for issue and payment. The results presented in Figure \ref{Fig.8}a and Figure \ref{Fig.8}b show that throughput drops and latency increases linearly with increasing the number of clients (and consequently, the total number of invoices). This is observed for all types of payments, either with or without split payment. It is therefore shown that the MTS is scalable for increasing transaction load when the increase is caused by adding more clients to the network.

\subsection{Findings of the scalability study}
The results of the scalability study have shown that the volume of data contained in an invoice has no discernible effect on the performance of the Smart Money system, therefore the design and implementation demonstrate scalability. This finding indicates that there is no restriction as to which businesses can participate in this scheme as sellers, with respect to the volume of goods that can be sold in a single transaction. For example, a shopping list in a supermarket is commonly larger compared to a shopping list from a clothing store. It also proves that MTS can be used not just for retail but also for wholesale transactions, where the quantity of goods exchanged can be much higher. Moreover, the Smart Money system can cope with increasing number of clients. The system can retain a certain level of transaction throughput, but the latency of each transaction increases with the number of clients, albeit linearly, therefore in a scalable manner. This shows that MTS is applicable to existing payment systems which have to support a very big number of endpoints (e.g. Point of Sales systems). The results also show that the Smart Money system scales well when the transaction load is increased, either when increasing the number of clients or when increasing the volume of transactions generated by each client. This is a requirement of modern payment systems which incorporate new forms of money such as CBDC, that will take up an even larger proportion of daily payments, as physical cash and cheques are being abandoned. Issuing an invoice is faster than paying for an invoice. This is caused by the extra computational overhead for the additional calculations and necessary checks e.g. check if it is an un-paid invoice, check if the funds required are available, check if this invoice can be paid ). 

Overall, the design has been demonstrated scalable within the simulation environment and the constraints of its hardware as described in section \ref{SMCordaDLTEnv}. 

\section{Conclusions}
As CBDCs are being presented and experimented with by central banks across the globe, it becomes apparent that the existing CBDC blueprints do not fully exploit the opportunities offered by DLT and smart contracts to aid central banks and OGDs in overcoming governance frictions and achieve their objectives. The research presented herein is based on an approach which integrates tax compliance to a CBDC without disrupting the existing money system. The proposed Smart Money CBDC, which is based on the R3 Corda DLT infrastructure, brings together for the first time the financial system and OGDs in a manner which is feasible, scalable, and practical for creating smart policies, such as automated tax collection from payments and controlled data access for policing. The inclusion of the programmable money concept in the Smart Money CBDC further amplifies the impact of the approach by extending the application to better control how social services are provided to citizens. These novel capabilities can only be realised with DLT-enabled smart contracts, which embody trust in multiparty agreements in a way that necessitates the revision of money’s standard definition and its role in the economy. Money can now act as a policy sensor and policy actuator by mobilising enhanced data contained within the payment messages. Smart Money, which is an enabler of Smart Policy, is the first practical CBDC application of a DLT which promises new capabilities for government with benefits for the economy and the society as a whole.

\section*{Acknowledgements}
We would like to acknowledge project partners the Bank of England, Office for National Statistics, Ministry of Justice and His Majesty's Revenue and Customs for contributing to advisory board meetings and project workshops.   

\section*{Funding}
This work was supported by the UK's Engineering and Physical Sciences Research Council [Grant number: \href{https://gow.epsrc.ukri.org/NGBOViewGrant.aspx?GrantRef=EP/P032001/1}{EP/P032001/1}]. Additional support was provided by EPSRC's Impact Acceleration Account (IAA).




 \bibliographystyle{elsarticle-num} 
 \bibliography{cas-refs}
\appendix

\section{Shopping lists}
\label{sec:sample:appendix}

\begin{lstlisting}[caption = Shopping list 1, label=listing1, basicstyle=\tiny]
[
  {
    "amount": 60626,
    "buyer": "Alice",
    "shoppingList": [
      {
        "item": "Groceries",
        "price": 3627,
        "quantity": 1,
        "vatRate": 0
      },
      {
        "item": "Energy",
        "price": 8040,
        "quantity": 1,
        "vatRate": 5
      },
      {
        "item": "Groceries",
        "price": 923,
        "quantity": 1,
        "vatRate": 0
      },
      {
        "item": "Books",
        "price": 9781,
        "quantity": 1,
        "vatRate": 0
      },
      {
        "item": "Groceries",
        "price": 4620,
        "quantity": 1,
        "vatRate": 0
      },
      {
        "item": "Children's Clothing",
        "price": 9159,
        "quantity": 2,
        "vatRate": 0
      },
      {
        "item": "Alcohol",
        "price": 9448,
        "quantity": 1,
        "vatRate": 20
      },
      {
        "item": "Adult Clothing",
        "price": 835,
        "quantity": 1,
        "vatRate": 20
      }
    ],
    "whoAmI": "MegaCompany"
  }
]
\end{lstlisting}

\begin{lstlisting}[caption = Shopping list 2, label=listing2, basicstyle=\tiny]
[
  {
    "amount": 26009,
    "buyer": "Alice",
    "shoppingList": [
      {
        "item": "Groceries",
        "price": 837,
        "quantity": 1,
        "vatRate": 0
      },
      {
        "item": "Groceries",
        "price": 1004,
        "quantity": 1,
        "vatRate": 0
      },
      {
        "item": "Energy",
        "price": 5090,
        "quantity": 1,
        "vatRate": 5
      },
      {
        "item": "Groceries",
        "price": 392,
        "quantity": 1,
        "vatRate": 0
      },
      {
        "item": "Books",
        "price": 3556,
        "quantity": 1,
        "vatRate": 0
      },
      {
        "item": "Groceries",
        "price": 8921,
        "quantity": 1,
        "vatRate": 0
      },
      {
        "item": "Children's Clothing",
        "price": 6209,
        "quantity": 2,
        "vatRate": 0
      }
    ],
    "whoAmI": "MegaCompany"
  }
]
\end{lstlisting}

\section{Pseudocodes for assertion tests}

\begin{algorithm}[!ht]
\caption{Pay invoice with sufficient funds}\label{alg:paysuff}
\begin{algorithmic}[1]
\Require $Totalamount$ $<$ $abalance$
\State $Shoppinglist = \{...\}$ \Comment{See Appendix, Listing \ref{listing1} for full list}
\State $Totalamount = Shoppinglist.amount$ 
\State $Netamount = Shoppinglist.netamount()$
\State $VAT = Shoppinglist.vatamount()$ 
\State $mcbalance = MegaCompany.getBalance()$
\State $abalance = Alice.getBalance()$
\State $vbalance = VATPayments.getBalance()$
\State $Inv = MegaCompany.issueOrder(Alice, Shoppinglist)$
\State $assert \{$ \Comment{Assertion Test 1}
\State \hspace{1cm} $Inv$ $\in$ $MegaCompany.getInvoices()$
\State \hspace{1cm} $Inv$ $\in$ $Alice.getInvoices()$
\State \hspace{1cm} $Inv$ $\not\in$ $VATPayments.getInvoices() \}$
\State $Alice.payInvoice(Inv)$
\State $assert \{$ \Comment{Assertion Test 2}
\State \hspace{1cm} $MegaCompany.getState(Inv)$ $==$ $isPaid$ 
\State \hspace{1cm} $Alice.getState(Inv)$ $==$ $isPaid$ 
\State \hspace{1cm} $VATPayments.getState(Inv)$ $==$ $isPaid \}$ 
\State $assert \{$ \Comment{Assertion Test 3}
\State \hspace{1cm} $MegaCompany.getAccountbalance(v)$ $==$ $mcbalance$ $+$ $Netamount$
\State \hspace{1cm} $Alice.getAccountbalance()$ $==$ $abalance$ $-$ $Totalamount$ 
\State \hspace{1cm} $VATPayments.getAccountbalance()$ $==$ $vbalance$ $+$ $VAT \}$ 
\end{algorithmic}
\end{algorithm}

\begin{algorithm}[!ht]
\caption{Pay invoice with insufficient funds}\label{alg:payinsuff}
\begin{algorithmic}[1]
\Require $Totalamount$ $>$ $abalance$
\State $Shoppinglist = \{...\}$ \Comment{See Appendix Listing \ref{listing1} for full list}
\State $Totalamount = Shoppinglist.amount$ 
\State $Netamount = Shoppinglist.netamount()$
\State $VAT = Shoppinglist.vatamount()$ 
\State $mcbalance = MegaCompany.getBalance()$
\State $abalance = Alice.getBalance()$
\State $vbalance = VATPayments.getBalance()$
\State $Inv = MegaCompany.issueOrder(Alice, Shoppinglist)$
\State $Alice.payInvoice(Inv)$
\State $assert \{$ \Comment{Assertion Test 4}
\State \hspace{1cm} $MegaCompany.getState(Inv)$ $!=$ $isPaid$ 
\State \hspace{1cm} $Alice.getState(Inv)$ $!=$ $isPaid$ 
\State \hspace{1cm} $VATPayments.getState(Inv)$ $!=$ $isPaid \}$ 
\State $assert \{$ \Comment{Assertion Test 5}
\State \hspace{1cm} $MegaCompany.getAccountbalance(v)$ $==$ $mcbalance$
\State \hspace{1cm} $Alice.getAccountbalance()$ $==$ $abalance$ 
\State \hspace{1cm} $VATPayments.getAccountbalance()$ $==$ $vbalance \}$ 
\end{algorithmic}
\end{algorithm}

\begin{algorithm}[!ht]
\caption{Pay invoice with goods within the \textit{allowed goods} list with tokens}\label{alg:paywithtokens}
\begin{algorithmic}[1]
\Require $Totalamount$ $<$ $atokenbalance$
\State $Shoppinglist = \{...\}$ \Comment{See Appendix, Listing \ref{listing2} for full list}
\State $Totalamount = Shoppinglist.amount$ 
\State $Netamount = Shoppinglist.netamount()$
\State $VAT = Shoppinglist.vatamount()$ 
\State $mcbalance = MegaCompany.getBalance()$
\State $atokenbalance = Alice.getTokenbalance()$
\State $vbalance = VATPayments.getBalance()$
\State $Inv = MegaCompany.issueOrder(Alice, Shoppinglist)$
\State $Alice.payInvoicewithTokens(Inv)$
\State $assert \{$ \Comment{Assertion Test 6}
\State \hspace{1cm} $MegaCompany.getState(Inv)$ $==$ $isPaid$ 
\State \hspace{1cm} $Alice.getState(Inv)$ $==$ $isPaid$ 
\State \hspace{1cm} $VATPayments.getState(Inv)$ $==$ $isPaid \}$ 
\State $assert \{$ \Comment{Assertion Test 7}
\State \hspace{1cm} $MegaCompany.getAccountbalance(v)$ $==$ $mcbalance$ $+$ $Netamount$
\State \hspace{1cm} $Alice.getAccountbalance()$ $==$ $atokenbalancebalance$ $-$ $Totalamount$
\State \hspace{1cm} $VATPayments.getAccountbalance()$ $==$ $vbalance$ $+$ $VAT \}$
\end{algorithmic}
\end{algorithm}

\begin{algorithm}[!ht]
\caption{Pay invoice with goods outside the \textit{allowed goods} list with tokens}\label{alg:paywithtokensnotallowed}
\begin{algorithmic}[1]
\Require $Totalamount$ $<$ $atokenbalance$
\State $Shoppinglist = \{...\}$ \Comment{See Appendix, Listing \ref{listing1} for full list}
\State $Totalamount = Shoppinglist.amount$ 
\State $Netamount = Shoppinglist.netamount()$
\State $VAT = Shoppinglist.vatamount()$ 
\State $mcbalance = MegaCompany.getBalance()$
\State $atokenbalance = Alice.getTokenbalance()$
\State $vbalance = VATPayments.getBalance()$
\State $Inv = MegaCompany.issueOrder(Alice, Shoppinglist)$
\State $Alice.payInvoicewithTokens(Inv)$
\State $assert \{$ \Comment{Assertion Test 8}
\State \hspace{1cm} $MegaCompany.getState(Inv)$ $!=$ $isPaid$ 
\State \hspace{1cm} $Alice.getState(Inv)$ $!=$ $isPaid$ 
\State \hspace{1cm} $VATPayments.getState(Inv)$ $!=$ $isPaid \}$ 
\State $assert \{$ \Comment{Assertion Test 9}
\State \hspace{1cm} $MegaCompany.getAccountbalance(v)$ $==$ $mcbalance$
\State \hspace{1cm} $Alice.getAccountbalance()$ $==$ $abalance$ 
\State \hspace{1cm} $VATPayments.getAccountbalance()$ $==$ $vbalance \}$ 
\end{algorithmic}
\end{algorithm}

\begin{algorithm}[!ht]
\caption{Smart Warrant}\label{alg:smartwar}
\begin{algorithmic}[1]
\State $mcDAR = VATInvestigator.DAR(MegaCompany,$
\State \hspace{0.5cm} $LegalAuthority)$ 
\State $SignedmcDAR = LegalAuthority.signDAR(mcDAR,$
\State $VATInvestigator)$
\item[]
\State $assert \{$ \Comment{Assertion Test 10}
\State \hspace{0.5cm} $SignedmcDAR$ $\in$ $VATInvestigator.getDARs()$
\State \hspace{0.5cm} $VATInvestigator.getState(SignedmcDAR)$ $==$ $unexecuted \}$
\State $mcdata$ $=$ $VATInvestigator.executeDAR(SignedmcDAR)$
\State $assert \{$ \Comment{Assertion Test 11}
\State \hspace{0.5cm} $VATInvestigator.getState(SignedmcDAR)$ $!=$ $unexecuted\}$
\State $mcinvoices$ $==$ $vaultQuery(MegaCompany, invoices)$
\State $assert \{mcinvoices == mcdata\}$ \Comment{Assertion Test 12}
\State $assert \{$ \Comment{Assertion Test 13}
\State \hspace{0.5cm} $NULL$ $==$ $vaultQuery(MegaCompany, invoices)\}$
\end{algorithmic}
\end{algorithm}





\end{document}